\documentclass{aa}
\usepackage{natbib}
\bibpunct{(}{)}{;}{a}{}{,}
\usepackage{graphicx} 

\newenvironment{equationarray}
{\arraycolsep 0.14 em
\begin {eqnarray}}
{\end {eqnarray}}

\newenvironment{equationarray*}
{\arraycolsep 0.14 em
\begin {eqnarray*}}
{\end {eqnarray*}}

\makeatletter
\def\gsim{\compoundrel>\over\sim}
\def\lsim{\compoundrel<\over\sim}
\def\compoundrel#1\over#2%
  {\mathpalette\compoundreL{{#1}\over{#2}}}
\def\compoundreL#1#2{\compoundREL#1#2}
\def\compoundREL#1#2\over#3%
  {\mathrel{\vcenter{\hbox{$\m@th \buildrel{#1#2}\over{#1#3}$}}}}
\makeatother

\newlength{\zero}
\newcommand{\wz}{\hspace{\zero}}
\newlength{\aste}
\newcommand{\wa}{\hspace{\aste} }
\newcommand{\R}{$ ^{\rm R} $}
\newcommand{\N}{$ ^{\rm N} $}
\newcommand{\A}{$ ^* $}
\newcommand{\lowrow}[1]{\smash{\lower 1.5 ex \hbox{#1}}}

\begin{document}

\title{Relativistic simulations of rotational core collapse. \\
II. Collapse dynamics and gravitational radiation}

\author{Harald Dimmelmeier \inst{1} 
\and Jos\'e A. Font \inst{1,2} 
\and Ewald M\"uller \inst{1}}

\offprints{Harald Dimmelmeier, \\ \email{harrydee@mpa-garching.mpg.de}}

\institute{Max-Planck-Institut f\"ur Astrophysik,
Karl-Schwarzschild-Str. 1, 85741 Garching, Germany
\and
Departamento de Astronom\'{\i}a y Astrof\'{\i}sica,
Universidad de Valencia, 46100 Burjassot (Valencia), Spain}

\date{Received date / Accepted date}


\abstract{ 
  We have performed hydrodynamic simulations of relativistic
  rotational supernova core collapse in axisymmetry and have computed
  the gravitational radiation emitted by such an event. The Einstein
  equations are formulated using the conformally flat metric
  approximation, and the corresponding hydrodynamic equations are
  written as a first-order flux-conservative hyperbolic
  system. Details of the methodology and of the numerical code have
  been given in an accompanying paper. We have simulated the evolution
  of 26 models in both Newtonian and relativistic gravity. The
  initial configurations are differentially rotating relativistic
  $ 4 / 3 $-polytropes in equilibrium which have a central density of
  $ 10^{10} {\rm\ g\ cm}^{-3}$. Collapse is initiated by decreasing
  the adiabatic index to some prescribed fixed value. The equation of
  state consists of a polytropic and a thermal part for a more
  realistic treatment of shock waves. Any microphysics like
  electron capture and neutrino transport is neglected.
  Our simulations show that the three different types of rotational
  supernova core collapse and gravitational waveforms identified in
  previous Newtonian simulations (regular collapse, multiple bounce
  collapse, and rapid collapse) are also present in relativistic
  gravity. However, rotational core collapse with multiple bounces is
  only possible in a much narrower parameter range in relativistic
  gravity. The relativistic models cover almost the same range of
  gravitational wave amplitudes
  ($ 4 \times 10^{-21} \le h^{\rm TT} \le 3 \times 10^{-20} $ for a
  source at a distance of 10~kpc) and
  frequencies ($ 60 {\rm\ Hz} \le \nu \le 1000 {\rm\ Hz} $) as the
  corresponding Newtonian ones. Averaged over all models, the total
  energy radiated in form of gravitational waves is
  $ 8.2 \times 10^{-8}\, M_{\odot} c^2 $
  in the relativistic case, and $ 3.6 \times 10^{-8}\, M_{\odot} c^2 $
  in the Newtonian case. For all collapse models which are of the
  same type in both Newtonian and relativistic gravity, the
  gravitational wave signal is of lower amplitude. If the collapse
  type changes, either weaker or stronger signals are found in the
  relativistic case. For a given model, relativistic gravity can
  cause a large increase of the characteristic signal frequency of up
  to a factor of five, which may have important consequences for the signal
  detection. Our study implies that the prospects for
  detection of gravitational wave signals from axisymmetric supernova
  rotational core collapse do not improve when taking into account
  relativistic gravity. The gravitational wave signals obtained in our
  study are within the sensitivity range of the first generation laser
  interferometer detectors if the source is located within the Local
  Group. An online catalogue containing the gravitational wave signal
  amplitudes and spectra of all our models is available at the URL
  {\tt http://www.mpa-garching.mpg.de/Hydro/hydro.html}.
  \keywords{Gravitation -- Gravitational waves -- Hydrodynamics --
    Stars: neutron -- Stars: rotation -- Supernovae: general}}

\authorrunning{H.~Dimmelmeier, J.A.~Font \& E.~M\"uller}
\titlerunning{Relativistic rotational core collapse. II}

\maketitle


\section{Introduction}
\label{sec:introduction}

At the end of their thermonuclear evolution, massive stars develop a
core composed of iron group nuclei (hence iron core) which becomes
dynamically unstable against gravitational collapse. The iron core
collapses to a neutron star or a black hole releasing gravitational
binding energy of the order
$ \sim 3 \times 10^{53} {\rm\ erg} \, (M / M_\odot)^2 (R / 10 {\rm\ km})^{-1}$,
which is sufficient to power a supernova explosion. If the core
collapse and/or the supernova explosion are nonspherical, part of the
liberated gravitational binding energy will be emitted in the form of
gravitational waves. Nonsphericity can be caused by the effects of
rotation, convection and anisotropic neutrino emission leading either
to a large-scale deviation from spherical symmetry or to small-scale
statistical mass-energy fluctuations (for a review, see e.g.,
\citet{mueller_98_a}).

According to present knowledge the energy radiated away in form of
gravitational waves does not exceed $ 10^{-6} M_\odot c^2 $ in rotational
core collapse \citep{mueller_82_a, finn_90_a, moenchmeyer_91_a,
bonazzola_93_a, yamada_95_a, zwerger_97_a, rampp_98_a, fryer_02_a} and
from convection or anisotropic neutrino emission in neutrino-driven
supernovae \citep{mueller_97_a}. The frequency of the emitted
radiation ranges from about a few Hz to a few kHz, and the
(dimensionless) signal amplitudes for a source located at a distance
of 10~Mpc (within the Virgo cluster) do not exceed $ \sim 10^{-22} $.
The smallness of the signals calls for accurate waveform templates to
alleviate the technical difficulties in extracting the signal from
noisy data \citep{pradier_01_a}. For the current generation of laser
interferometric detectors (GEO~600, LIGO, VIRGO, TAMA) such small
amplitudes imply that the prospects for detection of gravitational
waves from core collapse supernovae are limited to those events
occurring within the Local Group. However, if measured, a
gravitational wave signal does provide (as observations of neutrinos)
a direct diagnosis of the dynamics of the events.

Investigations of rotational core collapse are also important in the
context of the supernova explosion mechanism \citep{mueller_81_a,
bodenheimer_83_a, symbalisty_84_a, moenchmeyer_89_a, janka_89_a,
imshennik_92_a, yamada_94_a, fryer_00_a} and for the collapsar
scenario of gamma-ray bursts, where the collapse of a rotating
massive star leads to the formation of a Kerr black hole and a
relativistic jet \citep{mac_fadyen_01_a, aloy_00_a, wheeler_00_a}.

Numerous studies have addressed quite different aspects of rotational
core collapse at different levels of sophistication. Ideally, such
studies should incorporate a relativistic treatment of gravity and
hydrodynamics without symmetry restrictions, a detailed description of
the complex microphysics and of the neutrino transport, and should use
consistent initial models from evolutionary calculations of rotating
stars. The latter point is still a nagging problem (for a major step
forward in this context see \citet{heger_00_a}), as up to now all
studies, except that of \citet{fryer_02_a}, have relied on
parameterized initial models.

As the general relativistic corrections of the gravitational potential
during supernova core collapse, and for a neutron star, do not exceed
30\%, it is often argued that they can be neglected or only taken into
account approximately. While this approach may be justified to some
degree in the case of non-rotational core collapse, it becomes very
questionable when the core possesses a significant amount of angular
momentum. In the latter case, a relativistic treatment of gravity is
much more important, because the stabilizing effect of rotation is
counteracted by the destabilizing effect of the deeper relativistic
potential. This is reflected in the expression for the critical
adiabatic index $ \gamma_{\rm crit} $, below which a rigidly rotating
relativistic configuration is dynamically unstable against
pseudo-radial (linear) isentropic perturbations (see, e.g., Chap.~14
of \citet{tassoul_78_a}):
\begin{equation}
  \gamma_{\rm crit} = \frac{4}{3} - \frac{2}{9}
  \frac{\Omega^2 I}{|E_{\rm pot}|} + k \frac{R_{\rm S}}{R}.
  \label{eq:critical_adiabatic_index}
\end{equation}
Here $ \Omega $, $ I $, $ E_{\rm pot}$, $ R_{\rm S} $, and $ R $ are
the (constant) angular velocity, the moment of inertia about the
center of mass, the gravitational potential energy, the Schwarzschild
radius, and the radius of the star, respectively. The positive
constant $ k $ depends on the density distribution of the star.

Axisymmetric Newtonian hydrodynamic collapse and explosion simulations
using a realistic equation of state (EoS) and some treatment of weak
interaction processes have been performed by \citet{mueller_82_a},
\citet{bodenheimer_83_a}, \citet{symbalisty_84_a} and
\citet{bonazzola_93_a} neglecting neutrino transport, and by
\citet{moenchmeyer_89_a}, \citet{janka_89_a},
\citet{moenchmeyer_91_a}, \citet{imshennik_92_a}, \citet{fryer_00_a}
and \citet{fryer_02_a} employing some approximative description of
neutrino transport. In addition, \citet{finn_90_a},
\citet{yamada_94_a} and \citet{zwerger_97_a} have performed Newtonian
parameter studies of the axisymmetric collapse of rotating
polytropes. \citet{rampp_98_a} and \citet{brown_01_a} have extended
the work of \citet{zwerger_97_a} by relaxing the assumption of
axisymmetry.

\citet{wilson_79_a}, \citet{evans_86_a}, \citet{nakamura_81_a,
nakamura_83_a} and \citet{stark_85_a} pioneered investigations of
axisymmetric collapse of rotating configurations in full general
relativity (GR). \citet{wilson_79_a} computed neutron star bounces of
$ \gamma = 2 $ polytropes, while \citet{nakamura_81_a} (see also
\citet{nakamura_87_a}) simulated the formation of rotating black holes
resulting from the collapse of a $ 10 M_\odot $ ``core'' of a massive star
with different amounts of rotational energy and an initial central
density of $ 3 \times 10^{13} {\rm\ g\ cm}^{-3}
$. \citet{nakamura_83_a} (see also \citet{nakamura_87_a}) considered a
configuration consisting of a neutron star ($ M = 1.09 M_\odot $,
$ \rho_{\rm c} = 10^{15} {\rm\ g\ cm}^{-3} $) with an
accreted envelope of $ 0.81 M_\odot $, which was thought to mimic mass
fall-back in a supernova explosion. To this configuration he added
rotation and infall velocity, and simulated the evolution depending on
the prescribed rotation rates and rotation laws. In both scenarios the
EoS consisted of a relativistic degenerate lepton gas
($ \gamma = 4 / 3 $) at low densities
($ \rho \le \rho^\ast \equiv 3 \times 10^{14} {\rm\ g\ cm}^{-3} $, and
of a stiff ($ \gamma = 2 $) component at large densities
($ \rho > \rho^\ast $). \citet{stark_85_a} were the first to compute the
gravitational radiation from the relativistic collapse of a rotating
polytropic ($\gamma = 2$) star to a black hole. The initial model was
a spherically symmetric relativistic polytrope in equilibrium of mass
$ M $, central density $ 1.9 \times 10^{15} (M / M_\odot)^{-2} $, and radius
$ 6 G M / c^2 = 8.8 \times 10^5 M / M_\odot {\rm\ cm} $. Rotational
collapse was induced by lowering the pressure in the initial model by
a prescribed fraction, and by simultaneously adding an angular
momentum distribution approximating rigid-body
rotation. \citet{stark_85_a} found a low efficiency of gravitational
wave emission ($ E_{\rm rad\,tot} / M c^2 < 7 \times 10^{-4} $, where
$ E_{\rm rad\,tot} $ is the total energy radiated by graviational
waves), and that for sufficient rotation the star bounces and no black
hole forms.

After the work of \citet{stark_85_a} it took about 15 years before the
next simulations of GR rotational core collapse were published. This
was mainly caused by persistent numerical problems occurring in
axisymmetric GR simulations due to coordinate singularities at the
symmetry (=\,rotation) axis. These coordinate singularities hamper the
development of methods which guarantee accuracy and stability in long
term (covering many dynamical time scales) simulations.
\citet{alcubierre_00_a} proposed a method which does not suffer from
stability problems and where, in essence, Cartesian coordinates are
used even for axisymmetric systems. Using this method
\citet{shibata_00_a} investigated the effects of rotation on the
criterion for prompt adiabatic collapse of rigidly and differentially
rotating ($ \gamma = 2 $) polytropes to a black hole. Collapse of the
initial approximate (computed by assuming a conformally flat spatial
metric) equilibrium models was induced by a pressure
reduction. \citet{shibata_00_a} found that the criterion for black
hole formation depends strongly on the amount of angular momentum, but
only weakly on its (initial) distribution. He also studied the
effects of shock heating using a gamma-law EoS, and found that shock
heating is important in preventing prompt collapse to black holes in
case of large rotation rates.

\citet{hayashi_99_a} investigated the possibility of secular, i.e.\
quasi-static, core contraction from white dwarf to neutron star
densities using equilibrium sequences of rapidly rotating, general
relativistic compact stars with phenomenological equations of state.
They demonstrated that there is a possibility for the existence of
``fizzlers'' in the framework of GR, at least for the simplified equations
of state used in their study.

The above discussion shows that previous investigations in GR
rotational core collapse were mainly concerned with the question of
black hole formation under idealized conditions, but none of these
studies has really addressed the problem of supernova core collapse
which proceeds from white dwarf densities to neutron star densities,
involves core bounce, shock formation, and shock propagation. Exactly
this is the motivation for the present work. To this end, we apply the
numerical methodology presented in an accompanying paper
\citep[hereafter Paper~I]{dimmelmeier_02_a} and study the dynamics of
axisymmetric, relativistic rotational supernova core collapse and the
associated emission of gravitational radiation. In particular, we have
simulated the collapse of a comprehensive set of rotating relativistic
stellar models parameterized by the initial degree of differential
rotation, the initial rotation rate and the adiabatic indices at
subnuclear (supranuclear) matter densities. Thus, our study is a GR
extension of the previous work of \citet{zwerger_95_a} and
\citet{zwerger_97_a}, who computed the gravitational radiation from
rotational core collapse supernovae using Newtonian gravity. First
results from our study have already been published in a short
communication \citep{dimmelmeier_01_a}\footnote{Note that due to a
programming error, which we found only after the Letter was published,
the values of some quantities (e.g., the bounce densities and maximum
wave amplitudes) given in the Letter differ from those presented here,
but all qualitative results remain unchanged.}.

The organization of the paper is as follows:
Section~\ref{sec:model_assumptions} contains a brief repetition of
the assumptions made in our study (for more details see Paper~I). In
Section~\ref{sec:collapse_dynamics} we analyze in detail the collapse
dynamics and introduce the three different types of collapse and
gravitational waveforms we have identified.
Section~\ref{sec:gw_emission} is devoted to the gravitational wave
emission of the models, and to the prospects of detectability of
gravitational waves from core collapse supernovae. We conclude with a
summary in Section~\ref{sec:conclusions}. Additional information is
given in Appendix~\ref{sec:wave_extraction} where we discuss the
issue of gravitational wave extraction.


\section{Model assumptions}
\label{sec:model_assumptions}

The matter model obeys an ideal gas EoS with the pressure $ P $
consisting of a polytropic and a thermal part for a more realistic
treatment of the effects of shock waves \citep{janka_93_a,
zwerger_97_a}. Since we are mostly interested in the gravitational
radiation emission, which is controlled by the bulk motion of the
fluid, we neglect microphysics like electron capture and neutrino
transport. The initial configurations are differentially rotating
relativistic $ 4 / 3 $-polytropes in equilibrium \citep{komatsu_89_a,
komatsu_89_b, stergioulas_95_a} which are marginally stable, and
which have a central density
$ \rho_{\rm c\,ini} = 10^{10} {\rm\ g\ cm}^{-3}$. Collapse is
initiated by decreasing the adiabatic index (initially $ 4 / 3 $ in
all models) to some prescribed fixed value $ \gamma_1 $ (for more
details see Paper~I).

The initial models are determined by three parameters, which are also
used to name the models. The first parameter $ A $ is a length scale,
which specifies the degree of differential rotation (see Eq.~(33) of
Paper~I). The smaller the value of $ A $ the more differentially
rotating the model is. Model parameters A1, A2, A3 and A4 correspond
to $ A = 5 \times 10^9 {\rm\ cm} $, $ A = 10^8 {\rm\ cm} $,
$ A = 5 \times 10^7 {\rm\ cm} $, and $ A = 10^7 {\rm\ cm} $,
respectively. The second parameter is the initial rotation rate
$\beta_{\rm rot\,ini}$, which is given by the ratio of rotational
energy and the absolute value of the gravitational binding
energy. Model parameters B1, B2, B3, B4 and B5 correspond to
$ \beta_{\rm rot\,ini} = 0.25\%, 0.5\%, 0.9\%, 1.8\% $ and $ 4\% $,
respectively. The third parameter $ \gamma_1 $ is the adiabatic index
at subnuclear densities ($ \rho < \rho_{\rm nuc} $),
with model parameters G1, G2, G3, G4 and G5
corresponding to $ \gamma_1 = 1.325, 1.320, 1.310, 1.300 $ and
$ 1.280 $, respectively. The name of a simulated model is then given
by a combination of parameters from the three sets (e.g., A3B4G5).
We define the nuclear matter density
$ \rho_{\rm nuc} $ as $ 2.0 \times 10^{14} {\rm\ g\ cm}^{-3} $.

The adiabatic index at supranuclear densities
($ \rho \ge \rho_{\rm nuc} $) is fixed to $ \gamma_2 = 2.5 $, except
in one model which has also been run with $ \gamma_2 = 2.0 $ in order to
test the influence of a softer supranuclear EoS on the collapse dynamics.

In total, we have simulated the evolution of 26 models (see
Table~\ref{tab:models_summary}). In order to identify the relativistic
effects on the collapse dynamics, we have simulated these models also
in Newtonian gravity. Note that the Newtonian simulations are a subset
of those performed by \citet{zwerger_95_a} and \citet{zwerger_97_a}.

The Einstein equations are formulated using the so-called conformally
flat (CF) metric approximation \citep[conformal flatness condition --
CFC]{wilson_96_a} and the corresponding hydrodynamic equations are
formulated as a first-order flux-conservative hyperbolic system
\citep{banyuls_97_a}, well-adapted to numerical schemes based on
Riemann solvers. The applicability and quality of the CFC for
rotational core collapse has been discussed in detail in Paper~I,
where we demonstrated that its usage is appropriate.


\section{Collapse dynamics}
\label{sec:collapse_dynamics}


\subsection{Collapse and waveform types}
\label{subsec:collapse_and_waveform_types}

\citet{zwerger_95_a} first identified three distinct classes of core
collapse types. This classification is based on the form of the
gravitational wave signal. However, as the gravitational radiation
waveform is closely linked to the collapse dynamics, it also mirrors
the collapse behavior. Thus, the signal types can generally be used to
classify the collapse type, too.

In our relativistic simulations, each of the examined models belongs
to one of Zwerger's three collapse types, which he called type~I, II
and III. In the following we describe a representative model from each
class, discuss its characteristic properties and explain the physical
effects which lead to its collapse classification.

In general, the evolution of any supernova core collapse can be
divided into three phases (see, e.g., \citet{mueller_98_a}):

{\it Infall phase}: This phase covers the initial collapse of the core
from the onset of the gravitational instability triggered by the
sudden softening of the EoS due to the reduction of the subnuclear
adiabatic index. The inner part of the core, which collapses
homologously ($ v_r \sim r $), constitues the ``inner core'', while
the ``outer core'' is falling supersonically. Depending on the model
parameters the infall phase lasts between 30~ms and 100~ms.

{\it Bounce phase}: When the EoS stiffens because repulsive nuclear
forces become important at densities above nuclear density
$ \rho_{\rm nuc} = 2 \times 10^{14} {\rm\ g\ cm}^{-3} $, or when
centrifugal forces begin to dominante over gravitational attraction
due to angular momentum conservation and rotational spin-up, the inner
core decelerates on a timescale of about 1~ms. Because of its large
inertia and infall kinetic energy, the inner core does not come to
rest immediately. The core overshoots the equilibrium configuration it
will eventually approach, and bounces back, which results in the
formation of a shock wave at the outer edge of the inner core. During
core bounce the amplitude of the gravitational wave signal is largest.

{\it Ring-down phase, or re-expansion phase}: If centrifugal forces
remain sufficiently small until nuclear density is reached in the
center of the core, the bounce occurs at central densities slightly in
excess of nuclear density due to the stiffening of the EoS. In this
case the inner core rapidly settles down into a new equilibrium state,
which we will call the ``compact remnant''\footnote{Due to neutrino
cooling and other microphysical processes, which we do not consider in
our models, the compact remnant later evolves into a
neutron star. This may be accompanied by mass accretion due to
fall-back. We have not investigated this late time evolution of the
compact remnant.} in the following. While the shock wave propagates
outwards, the inner core oscillates for about 10\,ms with a
superposition of several damped eigenmodes with frequencies of about
$ 10^3 {\rm\ Hz} $ (ring-down). If the infall is moderately fast
(collapse timescale longer than about 50~ms), we will talk of a
regular collapse model giving rise to a type~I gravitational wave
signal according to the nomenclature of \citet{zwerger_97_a}. In case
of a very rapid infall (collapse timescale of about 30~ms), the core
plunges unimpeded by centrifugal forces deeply into the gravitational
potential well, with its central density significantly exceeding
nuclear density. In such rapid collapse models, the gravitational
waveform (type~III according to the nomenclature of
\citet{zwerger_97_a}) is qualitatively different from that of a
regular (type~I) collapse model (see below).

On the other hand, if core collapse is only or predominantly stopped
by centrifugal forces, the inner core experiences several distinct
sequences of infall, bounce, and re-expansion separated by up to
50~ms before it eventually settles down into an equilibrium state.
During each bounce an outward shock is generated. The multiple large
scale bounces occur because the deceleration by centrifugal forces is
less abrupt than that due to the stiffness of the supranuclear
EoS. The latter is characterized by an adiabatic index
$ \gamma_2 = 2.5 $ (see Section~\ref{sec:model_assumptions}), while
rotation acts like a $ \gamma = 5 / 3 $ gas according to the virial
theorem (see, e.g., \citet{tassoul_78_a}, Chapter~14). Multiple bounce
models produce type~II gravitational wave signals \citep{zwerger_97_a}
which consist of a distinct large amplitude peak for every bounce.

If the peak central density at bounce is close to (or exceeds)
nuclear matter density, and if the core rotates differentially and
rapidly, a mixture of a regular and multiple collapse type model can
occur. In these transition models of type~I/II the core re-expands
less than in a genuine multiple bounce model, but it still exhibits
distinct bounces and coherent large scale oscillations.


\subsubsection{Regular collapse -- type~I}
\label{subsec:regular_collapse}

\begin{figure}
  \resizebox{\hsize}{!}{\includegraphics{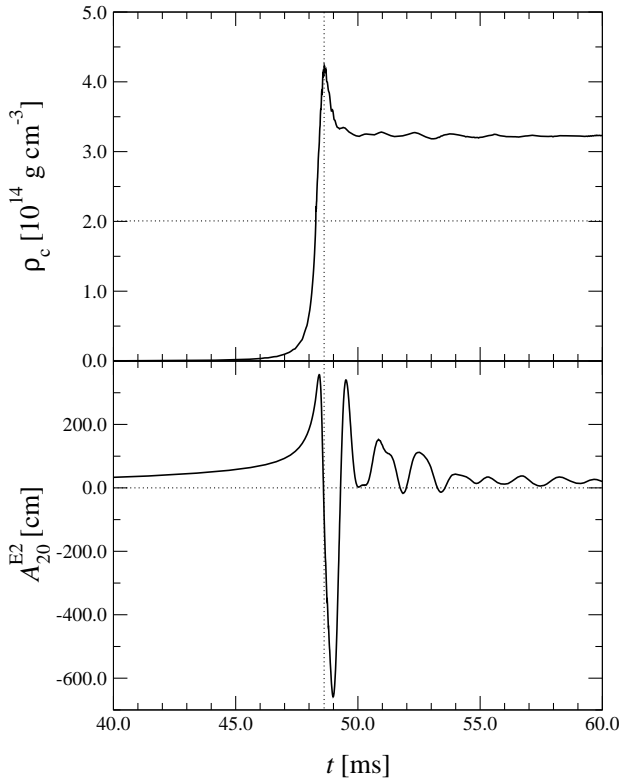}}
  \caption{Time evolution of the central density $ \rho_{\rm c} $
    (upper panel) and the gravitational wave signal amplitude
    $ A^{\rm E2}_{20} $ (lower panel) for the {\it regular collapse}
    model A1B3G3. The horizontal dotted line in the upper panel marks
    nuclear matter density $ \rho_{\rm nuc} $, and the vertical dotted
    line indicates the time of peak central density $ t_{\rm b} $.}
\label{fig:regular_collapse}
\end{figure}

As in the work of \citet{zwerger_97_a} we choose model A1B3G3 as a
representative model for a regular collapse.
Fig.~\ref{fig:regular_collapse} shows the time evolution of the
central density $ \rho_{\rm c} $ and the gravitational wave signal
amplitude $ A^{\rm E2}_{20}$; the definition of the wave amplitude
$ A^{\rm E2}_{20} $ and the description of the numerical wave
extraction technique are given in
Appendix~\ref{sec:wave_extraction}. The three phases of the collapse
are clearly identifiable in both panels. During the infall phase $
\rho_{\rm c} $ increases and exceeds nuclear matter density at
48.26\,ms. The peak value of the density is reached at the time of
bounce at $ t_{\rm b} = 48.63 {\rm\ ms} $ with
$ \rho_{\rm c\,b} = 4.23 \times 10^{14} {\rm\ g\ cm}^{-3} $, which is
about twice nuclear matter density. Subsequently, the core slightly
re-expands and rings down to an equilibrium state with a central
density of $ \rho_{\rm c\,f} \approx 1.5 \rho_{\rm nuc} $.

The gravitational wave amplitude (lower panel of
Fig.~\ref{fig:regular_collapse}) increases during the infall phase,
but has a negative peak value with
$ |A^{\rm E2}_{20}|_{\rm max} = 659 {\rm\ cm} $ at
$ t_{\rm gw} = 48.99 {\rm\ ms} > t_{\rm b} $. In all regular collapse
models the maximum signal amplitude is negative. For some models (not
in model A1B3G3) the signal amplitude exhibits a small local minimum
around the time of bounce, a feature which was identified and
discussed by \citet{zwerger_97_a}. The ring-down phase of the
pulsating inner core is directly reflected in the wave signal which
oscillates accordingly. As in the Newtonian runs of
\citet{zwerger_97_a}, we find that during the ring-down phase the
maxima of $ A^{\rm E2}_{20} $ are less damped than the minima, because
larger accelerations are encountered at the high density extrema of
the core.


\subsubsection{Multiple bounce collapse -- type~II}
\label{subsec:multiple_bounce_collapse}

\begin{figure}
  \resizebox{\hsize}{!}{\includegraphics{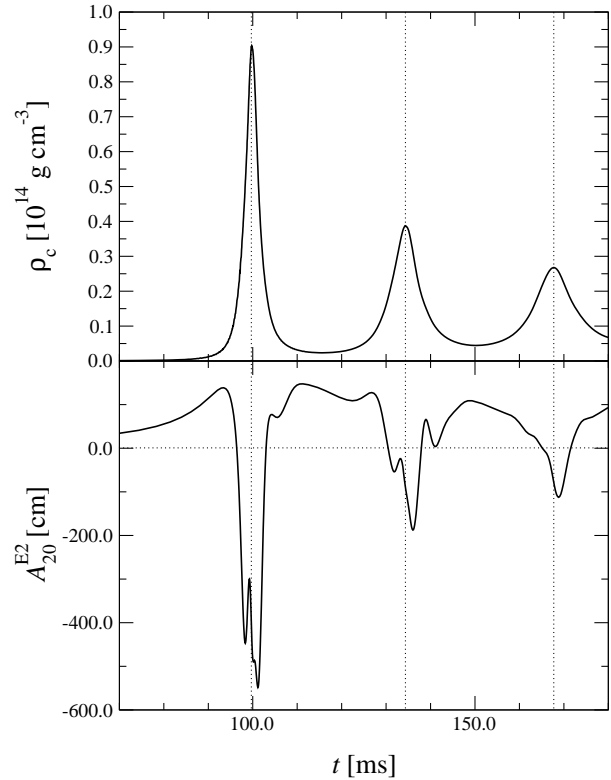}}
  \caption{Time evolution of the central density $ \rho_{\rm c} $
  (upper panel) and the gravitational wave signal amplitude
  $ A^{\rm E2}_{20} $ (lower panel) for the {\it multiple bounce
  collapse} model A2B4G1. The vertical dotted lines mark the times
  of peak central density for each bounce.}
\label{fig:multiple_bounce_collapse}
\end{figure}

For an adiabatic index $ \gamma_1 $ close to the initial value
$ \gamma_{\rm ini} = 4 / 3 $ (see Section~\ref{sec:model_assumptions}),
and for rapid and highly differential rotation, centrifugal forces can
halt the collapse at densities below nuclear matter density. Such
subnuclear bounces can indeed be observed in some of our collapse
models, e.g.\ model A2B4G1, for which the time evolution of the
central density $ \rho_{\rm c} $ and the gravitational wave signal
amplitude $ A^{\rm E2}_{20} $ are shown in
Fig.~\ref{fig:multiple_bounce_collapse}.

In both the central density evolution and the signal waveform distinct
extrema are discernable. The (negative) maximum of the signal
amplitude $ |A^{\rm E2}_{20}|_{\rm max} = 548 {\rm\ cm} $ is reached
at a time $ t_{\rm gw} = 101.16 {\rm\ ms} $, which is shortly after
the time $ t_{\rm b} = 99.78 {\rm\ ms} $ when the central density
reaches its first peak
$ \rho_{\rm c\,b} = 0.90 \times 10^{14} {\rm\ g\ cm}^{-3} \approx 0.45 \rho_{\rm nuc} $.
During the subsequent bounces the peaks of the central density are
significantly smaller than the peak value at first bounce. The
prominent peaks in the gravitational wave signal are clearly
associated with the distinct bounces visible in the density evolution.

We point out that model A1B3G1, which shows multiple bounces in
Newtonian gravity \citep{zwerger_97_a}, does not exhibit such a
behavior in the relativistic simulations. This change of collapse type
due to relativistic effects is discussed in
Section~\ref{subsec:change_of_dynamics}.

\begin{figure}
  \resizebox{\hsize}{!}{\includegraphics{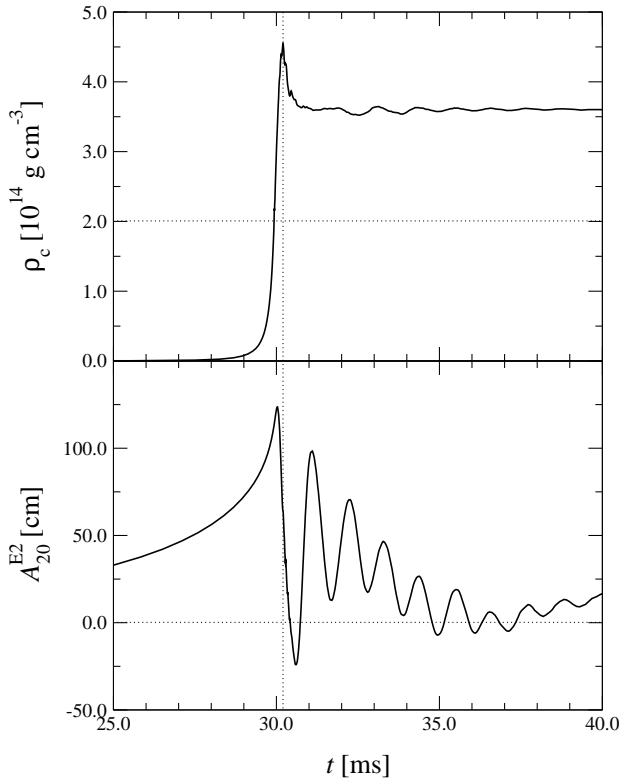}}
  \caption{Time evolution of the central density $ \rho_{\rm c} $
    (upper panel) and the gravitational wave signal amplitude
    $ A^{\rm E2}_{20} $ (lower panel) for the {\it rapid collapse}
    model A1B3G5. The horizontal dotted line in the upper panel marks
    nuclear matter density $ \rho_{\rm nuc} $, and the vertical dotted
    line indicates the time of peak central density
    $ t_{\rm b} $. Note the qualitative difference in the waveform
    compared to the regular collapse (type~I) model shown in
    Fig.~\ref{fig:regular_collapse}.}
\label{fig:rapid_collapse}
\end{figure}


\subsubsection{Rapid collapse -- type~III}
\label{subsec:rapid_collapse}

If the core collapses very rapidly due to values of $ \gamma_1 $ which
are much smaller than $4/3$, the qualitative behavior of the density
evolution is very similar to that of the regular (type~I) collapse
model, except that the bounce and equilibrium densities are slightly
higher, and the post-bounce pulsations of the inner core are even more
strongly suppressed. For example, in model A1B3G5
(Fig.~\ref{fig:rapid_collapse}), which is a typical rapid collapse
model, the values of the peak and the final central density are
$ \rho_{\rm c\,b} = 4.55 \times 10^{14} {\rm\ g\ cm}^{-3} \approx 2.5 \rho_{\rm nuc} $
(reached at $ t_{\rm b} = 30.20 {\rm\ ms} $), and
$ \rho_{\rm c\,f} \approx 1.8 \rho_{\rm nuc} $.

On the other hand, the shape of the gravitational wave signal
amplitude is qualitatively different compared to type~I models. This
is shown in the lower panel of Fig.~\ref{fig:rapid_collapse}, where
the time evolution of $ A^{\rm E2}_{20} $ is plotted. The maximum signal
amplitude $ |A^{\rm E2}_{20}|_{\rm max} = 124 {\rm\ cm} $ at
$ t_{\rm gw} = 30.03 {\rm\ ms} $ is now assigned to the first positive
peak of the waveform contrary to the signal waveform of the regular
collapse models. In the case of rapid collapse, the negative peak is almost
entirely suppressed and the overall amplitude of the signal is
significantly smaller. There is no clear feature in the waveform
allowing one to identify the time of bounce $ t_{\rm b} $.

For a given initial model, the transition between type~I and
type~III collapse models occurs gradually with decreasing values of $
\gamma_1 $. In the gravitational wave signal this is reflected by a
decrease of the negative main peak and an increase of the first
positive peak \citep{zwerger_97_a}.


\subsubsection{Additional remarks}
\label{subsec:collapse_type_remarks}

In the above models the central density $ \rho_{\rm c} $ coincides
with $ \rho_{\rm max} $, the maximum density of the model at any
time. However, very rapidly and differentially rotating models, like
all `A4' models, either already possess a toroidal density distribution
initially, or develop one during their evolution. In such models the
maximum density is attained at some off-center location, and
$ \rho_{\rm c} < \rho_{\rm max} $ holds. Nevertheless, the time
evolution of $ \rho_{\rm max} $ can be used to classify the collapse
type according to the criteria discussed above.

Our results show that relativistic simulations of rotational core
collapse exhibit the same qualitative collapse dynamics and allow for
the same classification as Newtonian simulations \citep{zwerger_95_a,
zwerger_97_a}. However, there exist important differences in other
aspects of the collapse arising from relativistic effects. These are
discussed in the subsequent sections.

\begin{table*}
  \centering
  \caption{Summary of important quantities for all models in
    relativistic (R) and Newtonian (N) gravity. $ t_{\rm b} $ is the
    time of bounce, $ \rho_{\rm max\,b} $ is the maximum density at
    bounce, $ |A^{\rm E2}_{20}|_{\rm max} $ is the maximum
    gravitational wave amplitude, $ \nu_{\rm max} $ is the frequency
    of maximum spectral energy density, $ \beta_{\rm rot\,max} $ is
    the maximum rotation rate, $ \rho_{\rm max\,f} $ is the maximum
    density after ring-down, and Type specifies the collapse type
    (for a definition see text). If the maximum density is located
    off-center, the respective density value is marked with an
    asterisk. For models of type~II or I/II, $ \rho_{\rm max\,f} $
    cannot be determined.}
  \setlength{\tabcolsep}{0.095 cm}
  \renewcommand{\arraystretch}{0.8}
  \begin{tabular}{lccccccccccr@{~~}}
    \hline
    ~~Model \rule{0 em}{1.6 em} &
    $ t_{\rm b} $ &
    $ \rho_{\rm max\,b} $ &
    $ \displaystyle \frac{\rho_{\rm max\,b}^{\rm R}}{\rho_{\rm max\,b}^{\rm N}} - 1 $ &
    $ |A^{\rm E2}_{20}|_{\rm max} $ &
    $ \displaystyle \frac{|A^{\rm E2}_{20}|_{\rm max}^{\rm R}}{|A^{\rm E2}_{20}|_{\rm max}^{\rm N}} - 1 $ &
    $ \nu_{\rm max} $ &
    $ \displaystyle \frac{\nu_{\rm max}^{\rm R}}{\mu_{\rm max}^{\rm N}} - 1 $ &
    $ \beta_{\rm rot\,max} $ &
    $ \displaystyle \frac{\beta_{\rm rot\,max}^{\rm R}}{\beta_{\rm rot\,max}^{\rm N}} - 1 $ &
    $ \rho_{\rm max\,f} $ &
    \multicolumn{1}{c}{Type} \\
    &
    [ms] &
    $ \left[ \displaystyle 10^{14} \frac{\rm g}{{\rm cm}^3} \right] $ &
    [\%] &
    [cm] &
    [\%] &
    [Hz] &
    [\%] &
    [\%] &
    [\%] &
    $ \left[ \displaystyle 10^{14} \frac{\rm g}{{\rm cm}^3} \right]_{} $ \\
    \hline
    A1B1G1\R  & 91.17 & 
      4.78\wa & \lowrow{$  \wz+48$} & \wz943 & \lowrow{$  -44$} &
    \wz623    & \lowrow{$  \wz+21$} & \wz5.5 & \lowrow{$\wz+6$} &
       3.5\wa & \rule{0 em}{0.9 em} I    \\
    A1B1G1\N  & 92.23 &
      3.22\wa &                    &   1698 & &
    \wz515    &                    & \wz5.2 & &
       ---\wa & I/II \\
    \hline
    A1B2G1\R  & 91.92 & 
      4.45\wa & \lowrow{$  \wz+54$} &   1803 & \lowrow{$  -23$} &
    \wz930    & \lowrow{$    +109$} & \wz8.9 & \lowrow{$\wz+9$} &
       3.4\wa & \rule{0 em}{0.9 em} I    \\
    A1B2G1\N  & 93.10 & 
      2.89\wa &                    &   2355 & &
    \wz446    &                    & \wz8.2 & &
       ---\wa & II   \\
    \hline
    A1B3G1\R  & 93.25 & 
      4.07\wa & \lowrow{$  \wz+88$} &   2679 & \lowrow{$  +84$} &
      1053    & \lowrow{$    +446$} &   12.6 & \lowrow{$  +18$} &
       3.2\wa & \rule{0 em}{0.9 em} I    \\
    A1B3G1\N  & 94.70 & 
      2.17\wa &                    &   1463 & &
    \wz193    &                    &   10.7 & &
       ---\wa & II   \\
    \hline
    A1B3G2\R  & 69.45 & 
      4.23\wa & \lowrow{$  \wz+45$} &   1718 & \lowrow{$  -18$} &
      1018    & \lowrow{$    +127$} &   11.9 & \lowrow{$  +12$} &
       3.2\wa & \rule{0 em}{0.9 em} I    \\
    A1B3G2\N  & 69.87 & 
      2.83\wa &                    &   2084 & &
    \wz448    &                    &   10.6 & &
       ---\wa & II   \\
    \hline
    A1B3G3\R  & 48.63 & 
      4.23\wa & \lowrow{$  \wz+25$} & \wz659 & \lowrow{$  -32$} &
    \wz702    & \lowrow{$  \wz+22$} &   10.6 & \lowrow{$  +33$} &
       3.2\wa & \rule{0 em}{0.9 em} I    \\
    A1B3G3\N  & 48.57 & 
      3.38\wa &                    & \wz976 & &
    \wz576    &                    & \wz8.0 & &
       2.5\wa & I    \\
    \hline
    A1B3G5\R  & 30.20 & 
      4.55\wa & \lowrow{$\wz\wz+7$} & \wz124 & \lowrow{$\wz-5$} &
    \wz904    & \lowrow{$\wz\wz+6$} & \wz5.1 & \lowrow{$  +50$} &
       3.6\wa & \rule{0 em}{0.9 em} III  \\
    A1B3G5\N  & 29.98 & 
      4.26\wa &                    & \wz131 & &
    \wz852    &                    & \wz3.4 & &
       3.0\wa & III  \\
    \hline \hline
    A2B4G1\R  & 99.78 & 
      0.90\wa & \lowrow{$    +718$} & \wz548 & \lowrow{$  -16$} &
  \wz\wz86    & \lowrow{$  \wz+41$} &   17.3 & \lowrow{$  +47$} &
       ---\wa & \rule{0 em}{0.9 em} II   \\
    A2B4G1\N  & 99.71 & 
      0.11\wa &                    & \wz652 & &
  \wz\wz61    &                    &   11.8 & &
       ---\wa & II   \\
    \hline \hline
    A3B1G1\R  & 91.63 & 
      4.61\wa & \lowrow{$  \wz+54$} &   1965 & \lowrow{$  -22$} &
      1044    & \lowrow{$    +117$} & \wz9.1 & \lowrow{$  +21$} &
       3.5\wa & \rule{0 em}{0.9 em} I    \\
    A3B1G1\N  & 92.72 & 
      3.00\wa &                    &   2510 & &
    \wz482    &                    & \wz7.5 & &
       ---\wa & II   \\
    \hline
    A3B2G1\R  & 92.86 & 
      4.24\wa & \lowrow{$  \wz+86$} &   2896 & \lowrow{$  +69$} &
      1133    & \lowrow{$    +148$} &   13.8 & \lowrow{$  +25$} &
       3.1\wa & \rule{0 em}{0.9 em} I    \\
    A3B2G1\N  & 94.32 & 
      2.28\wa &                    &   1711 & &
    \wz456    &                    &   11.0 & &
       ---\wa & II   \\
    \hline
    A3B2G2\R  & 69.53 & 
      4.10\wa & \lowrow{$  \wz+53$} &   2183 & \lowrow{$\wz-9$} &
      1128    & \lowrow{$    +128$} &   13.9 & \lowrow{$  +19$} &
       3.1\wa & \rule{0 em}{0.9 em} I    \\
    A3B2G2\N  & 69.98 & 
      2.68\wa &                    &   2407 & &
    \wz494    &                    &   11.7 & &
       ---\wa & II   \\
    \hline
    A3B2G4$ _{\rm soft}^{\rm R} $ & 39.38 & 
      6.03\wa & \lowrow{$  \wz+36$} & \wz618 & \lowrow{$  -21$} &
    \wz925    & \lowrow{$  \wz+23$} &   13.6 & \lowrow{$  +43$} &
       4.1\wa & \rule{0 em}{0.9 em} I    \\
    A3B2G4$ _{\rm soft}^{\rm N} $ & 39.14 & 
      4.45\wa &                    &  \wz781 & &
    \wz755    &                    &  \wz9.5 & &
       2.7\wa & I    \\
    \hline
    A3B2G4\R  & 39.34 & 
      4.05\wa & \lowrow{$  \wz+19$} & \wz517 & \lowrow{$  -26$} &
    \wz838    & \lowrow{$  \wz+14$} &   13.0 & \lowrow{$  +44$} &
       3.2\wa & \rule{0 em}{0.9 em} I    \\
    A3B2G4\N  & 39.07 & 
      3.41\wa &                    & \wz703 & &
    \wz737    &                    & \wz9.0 & &
       2.5\wa & I    \\
    \hline
    A3B3G1\R  & 95.26 & 
      3.49\wa & \lowrow{$    +642$} & \wz982 & \lowrow{$  -10$} &
    \wz827    & \lowrow{$    +347$} &   20.1 & \lowrow{$  +75$} &
       ---\wa & \rule{0 em}{0.9 em} I/II   \\
    A3B3G1\N  & 96.57 & 
      0.47\wa &                    &   1087 & &
    \wz185    &                    &   11.5 & &
       ---\wa & II   \\
    \hline
    A3B3G2\R  & 71.28 & 
      3.58\wa & \lowrow{$    +225$} &   1353 & \lowrow{$\wz-5$} &
    \wz846    & \lowrow{$    +209$} &   20.4 & \lowrow{$  +49$} &
       ---\wa & \rule{0 em}{0.9 em} I/II \\
    A3B3G2\N  & 71.77 & 
      1.10\wa &                    &   1420 & &
    \wz274    &                    &   13.7 & &
       ---\wa & II   \\
    \hline
    A3B3G3\R  & 49.73 & 
      3.35\wa & \lowrow{$  \wz+41$} &   1279 & \lowrow{$  -14$} &
    \wz895    & \lowrow{$    +147$} &   20.3 & \lowrow{$  +48$} &
       ---\wa & \rule{0 em}{0.9 em} I/II \\
    A3B3G3\N  & 49.64 & 
      2.38\A  &                    &   1480 & &
    \wz363    &                    &   15.7 & &
       ---\wa & I/II \\
    \hline
    A3B3G5\R  & 30.65 & 
      3.75\wa & \lowrow{$\wz\wz+9$} & \wz256 & \lowrow{$\wz-3$} &
      1074    & \lowrow{$  \wz+14$} &   13.7 & \lowrow{$  +47$} &
       3.4\wa & \rule{0 em}{0.9 em} III  \\
    A3B3G5\N  & 30.36 & 
      3.45\A  &                    & \wz263 & &
    \wz964    &                    & \wz9.3 & &
       2.3\A  & III  \\
    \hline
    A3B4G2\R  & 74.66 & 
      0.79\wa & \lowrow{$    +394$} & \wz594 & \lowrow{$  -34$} &
    \wz101    & \lowrow{$  \wz+19$} &   20.7 & \lowrow{$  +41$} &
       ---\wa & \rule{0 em}{0.9 em} II   \\
    A3B4G2\N  & 73.99 & 
      0.16\A  &                    & \wz894 & &
  \wz\wz85    &                    &   14.7 & &
       ---\wa & II   \\
    \hline
    A3B5G4\R  & 44.46 & 
      0.30\A  & \lowrow{$    +100$} & \wz487 & \lowrow{$\wz-8$} &
    \wz103    & \lowrow{$\wz\wz-9$} &   28.8 & \lowrow{$  +39$} &
       ---\wa & \rule{0 em}{0.9 em} III  \\
    A3B5G4\N  & 44.31 & 
      0.15\A  &                    & \wz528 & &
    \wz113    &                    &   20.7 & &
       ---\wa & III  \\
    \hline \hline
    A4B1G1\R  & 90.77 & 
      4.93\A  & \lowrow{$  \wz+69$} &   2520 & \lowrow{$  +27$} &
      1309    & \lowrow{$    +126$} &   10.5 & \lowrow{$  +42$} &
       3.5\A  & \rule{0 em}{0.9 em} I    \\
    A4B1G1\N  & 91.77 & 
      2.92\A  &                    &   1992 & &
    \wz580    &                    & \wz7.4 & &
       ---\wa & II   \\
    \hline
    A4B1G2\R  & 68.39 & 
      4.72\A  & \lowrow{$  \wz+53$} &   2220 & \lowrow{$\wz+9$} &
      1268    & \lowrow{$  \wz+34$} &   12.6 & \lowrow{$  +48$} &
       3.3\A  & \rule{0 em}{0.9 em} I    \\
    A4B1G2\N  & 68.69 & 
      3.09\A  &                    &   2034 & &
    \wz946    &                    & \wz8.5 & &
       ---\wa & II   \\
    \hline
    A4B2G2\R  & 68.61 & 
      4.79\A  & \lowrow{$  \wz+85$} &   3393 & \lowrow{$  +59$} &
      1346    & \lowrow{$    +189$} &   18.4 & \lowrow{$  +46$} &
       3.1\A  & \rule{0 em}{0.9 em} I    \\
    A4B2G2\N  & 69.02 & 
      2.59\A  &                    &   2132 & &
    \wz466    &                    &   12.6 & &
       ---\wa & II   \\
    \hline
    A4B2G3\R  & 48.87 & 
      4.37\A  & \lowrow{$  \wz+53$} &   2535 & \lowrow{$  +14$} &
      1336    & \lowrow{$    +261$} &   24.0 & \lowrow{$  +55$} &
       2.9\A  & \rule{0 em}{0.9 em} I    \\
    A4B2G3\N  & 48.74 & 
      2.85\A  &                    &   2217 & &
    \wz370    &                    &   15.5 & &
       ---\wa & I/II \\
    \hline
    A4B4G4\R  & 40.29 & 
      2.18\A  & \lowrow{$    +199$} &   1245 & \lowrow{$  -29$} &
    \wz155    & \lowrow{$  \wz-31$} &   31.8 & \lowrow{$  +41$} &
       ---\wa & \rule{0 em}{0.9 em} I/II \\
    A4B4G4\N  & 39.66 & 
      0.73\A  &                    &   1748 & &
    \wz226    &                    &   22.6 & &
       ---\wa & I/II \\
    \hline
    A4B4G5\R  & 32.32 & 
      3.08\A  & \lowrow{$  \wz+37$} &   1033 & \lowrow{$  -52$} &
    \wz227    & \lowrow{$\wz\wz+5$} &   43.2 & \lowrow{$  +54$} &
       ---\wa & \rule{0 em}{0.9 em} I/II \\
    A4B4G5\N  & 31.91 & 
      2.25\A  &                    &   2149 & &
    \wz217    &                    &   28.0 & &
       ---\wa & I/II \\
    \hline
    A4B5G4\R  & 38.26 & 
      0.98\A  & \lowrow{$    +158$} &   1651 & \lowrow{$  -44$} &
  \wz\wz90    & \lowrow{$  \wz+18$} &   38.4 & \lowrow{$  +38$} &
       ---\wa & \rule{0 em}{0.9 em} I/II \\
    A4B5G4\N  & 37.29 & 
      0.38\A  &                    &   2965 & &
  \wz\wz76    &                    &   27.9 & &
       ---\wa & I/II \\
    \hline
    A4B5G5\R  & 31.39 & 
      3.24\A & \lowrow{$  \wz+22$} &   2356 & \lowrow{$  -57$} &
    \wz106    & \lowrow{$  \wz-33$} &   48.6 & \lowrow{$  +28$} &
       ---\wa & \rule{0 em}{0.9 em} I/II \\
    A4B5G5\N  & 30.82 & 
      2.65\A  &                    &   5444 & &
    \wz159    &                    &   38.1 & &
       ---\wa & I/II \\
    \hline
  \end{tabular}
  \label{tab:models_summary}
\end{table*}


\subsection{Compactness of the core remnant}
\label{subsec:core_compactness}

\begin{figure}
  \resizebox{\hsize}{!}{\includegraphics{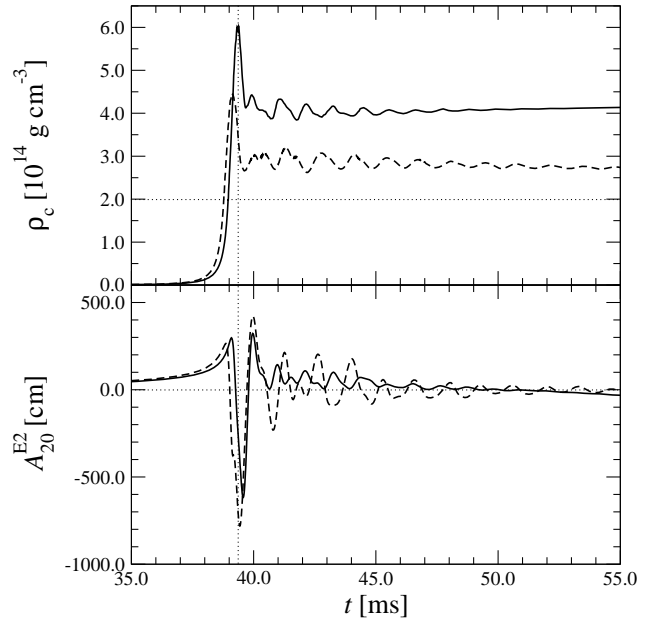}}
  \caption{Time evolution of the central density $ \rho_{\rm c} $
    (upper panel) and the gravitational wave signal amplitude
    $ A^{\rm E2}_{20} $ (lower panel) in the relativistic (solid
    lines) and Newtonian (dashed lines) simulation of the regular
    collapse model A3B2G4$ _{\rm soft} $. The horizontal dotted line
    in the upper panel marks nuclear matter density
    $ \rho_{\rm nuc} $. The vertical dotted line indicates the time of
    peak central density $ t_{\rm b} $.}
  \label{fig:comparison_regular_collapse}
\end{figure}

\begin{figure}
  \resizebox{\hsize}{!}{\includegraphics{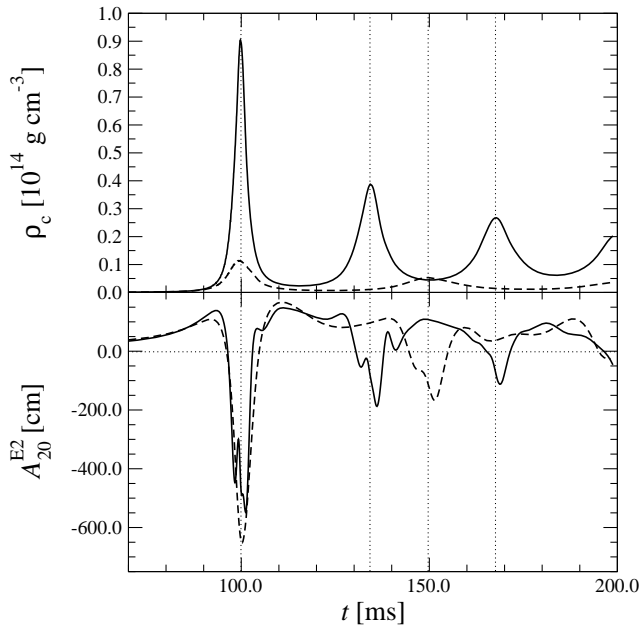}}
  \caption{Same as Fig.~\ref{fig:comparison_regular_collapse}, but
    for the multiple bounce collapse model A2B4G1. The vertical dotted
    lines mark the times of peak central density $ t_{\rm b} $ for
    each bounce.}
\label{fig:comparison_multiple_bounce_collapse}
\end{figure}

For all models the central density at bounce $ \rho_{\rm c\,b} $ (for
all A4 models and model A3B5G4 the off-center maximum density
$ \rho_{\rm max\,b} $) is larger in the relativistic simulation than in
the corresponding Newtonian one, the relative increase reaching
$ \sim 700\% $ in special cases (see
Table~\ref{tab:models_summary}). This also holds for the maximum
density of the compact remnant after ring-down in case of instant
formation of a stable equilibrium state (types~I and III). This
generic property is illustrated in the upper panel of
Fig.~\ref{fig:comparison_regular_collapse} and
Fig.~\ref{fig:comparison_multiple_bounce_collapse} for the type~I
model A3B2G4$ _{\rm soft} $ (where relativistic effects are
particularly large due to its soft EoS and thus high central density)
and the type~II model A2B4G1, respectively.

When comparing models of the same collapse type in Newtonian and
relativistic gravity, we find that although all relativistic models
reach higher central and/or maximum densities and, on average, larger
infall and rotation velocities, only six models (see
Table~\ref{tab:models_summary}) have also larger maximum signal
amplitudes than their Newtonian counterparts. The maximum
signal amplitudes of all other relativistic models
are (up to 57\%) smaller than those of the corresponding Newtonian ones
(Table~\ref{tab:models_summary}). As already discussed in
\citet{dimmelmeier_01_a}, the reduced maximum signal strength can be
explained by the fact that the amplitude, which is calculated
using the quadrupole formula (see Appendix~\ref{sec:wave_extraction}),
is determined by the {\em bulk} motion of the core rather than just by
the motion of the densest mass shells, strictly speaking by the second
time derivative of the quadrupole moment. Therefore, a core which is
more condensed in the center can give rise to a {\em smaller}
gravitational wave signal than a core which is less centrally
condensed, but which is denser and moves faster in its outer
regions. This is reflected by the weight factor $ r^2 $ in the
integrand of the quadrupole formula
(\ref{eq:standard_quadrupole_formula}). Due to this factor matter at
moderately large $ r $ contributes significantly to the gravitational
wave signal amplitude, while at very large radii (outside the inner
core) the decrease in density more than compensates the increase of
the weight factor.

\begin{figure}
  \resizebox{\hsize}{!}{\includegraphics{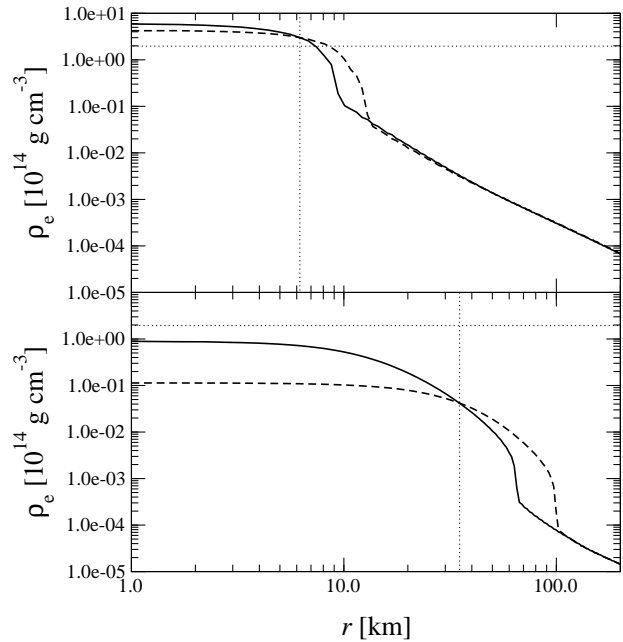}}
  \caption{Radial density profiles $ \rho_{\rm e} $ along the equator
    near bounce time for model A3B2G4$ _{\rm soft} $ (upper panel) and
    model A2B4G1 (lower panel). In both models, the density of the
    relativistic simulation (solid line) is larger than that of the
    Newtonian simulation (dashed line) in the central region. At
    $ r \approx 6 {\rm\ km} $ (upper panel) and $ r \approx 35 {\rm\ km} $
    (lower panel) the density of the relativistic model drops below
    that of the Newtonian one. The radii of the density crossing are
    indicated by the vertical dotted lines, and the horizontal dotted
    lines mark nuclear matter density $ \rho_{\rm nuc} $.}
  \label{fig:compact_core_density_profile}
\end{figure}

\begin{figure}
  \resizebox{\hsize}{!}{\includegraphics{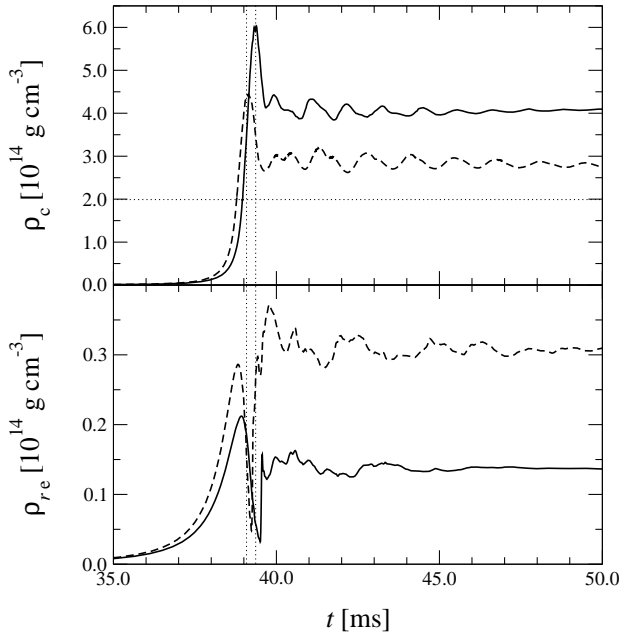}}
  \caption{Time evolution of the density at different radii in the
    relativistic (solid lines) and Newtonian (dashed lines) simulation
    of model A3B2G4$ _{\rm soft}$. Whereas the central density (upper
    panel) is larger in relativistic gravity, the (equatorial) density
    of the Newtonian model exceeds that of the relativistic model at a
    radius of $ r = 12.9 {\rm\ km} $ (lower panel). The vertical
    dotted lines indicate the time of bounce $ t_{\rm b} $, and the
    horizontal dotted line in the upper plot indicates nuclear matter
    density $ \rho_{\rm nuc} $. Note the different vertical scale of
    the two plots.}
  \label{fig:compact_core_density_evolution_regular_collapse}
\end{figure}

The upper panel of Fig.~\ref{fig:compact_core_density_profile} shows
the equatorial density profiles in the inner core of model
A3B2G4$_{\rm soft} $ for both the relativistic (solid line) and the
Newtonian (dashed line) simulation at bounce. As already seen in the
central density evolution
(Fig.~\ref{fig:comparison_regular_collapse}), the simulation in
relativistic gravity yields up to 36\% larger densities in the central
regions. However, at radii $ r \gsim 5 {\rm\ km} $ the density in the
Newtonian model exceeds that of the relativistic one. This ``density
crossing'', which takes place at different radii for different polar
angles $ \theta $, can be observed in all the models we have
investigated, at all times $ t > t_{\rm b} $ and at all polar
angles. The surface of density crossing is located inside the
compact remnant for type~I and type~III models, and in the case of
type~II multiple bounce models in the coherently re-expanding and
contracting inner core, which has not yet settled into a compact
remnant at the end of the simulation. The effect is persistent in time
as demonstrated in
Fig.~\ref{fig:compact_core_density_evolution_regular_collapse}, which
shows the time evolution of the density for model
A3B2G4$ _{\rm soft} $ at two different radial locations in the
equatorial plane. After bounce, the central density $ \rho_{\rm c} $
of the relativistic configuration is always larger than the
corresponding Newtonian one (upper panel). However, at larger radii,
but still inside the compact remnant (lower panel), the density of
the relativistic model is smaller by up to 50\%.

The analysis of the multiple bounce model A2B4G1, which yields a
smaller signal amplitude in relativistic gravity (see lower panel of
Fig.~\ref{fig:comparison_multiple_bounce_collapse}), leads to a similar
result as the regular bounce model A3B2G4$ _{\rm soft}$. In the
relativistic case, the central density at bounce is more than 8 times
larger than in the Newtonian one. As in model A3B2G4$ _{\rm soft}$,
the equatorial density profiles cross (lower panel of
Fig.~\ref{fig:compact_core_density_profile}), but now the crossing
occurs at a much larger radius $ r \approx 35 {\rm\ km} $. Hence, the
region where the relativistic model is denser is considerably more
extended. Inside the crossing radius, the cores of both the
relativistic and the Newtonian model oscillate coherently. The density
of the relativistic model is considerably higher in this region, and
the matter experiences stronger accelerations, which causes narrower
density peaks and shorter time intervals between two consecutive
bounces (Fig.~\ref{fig:comparison_multiple_bounce_collapse}).

\begin{figure}
  \resizebox{\hsize}{!}{\includegraphics{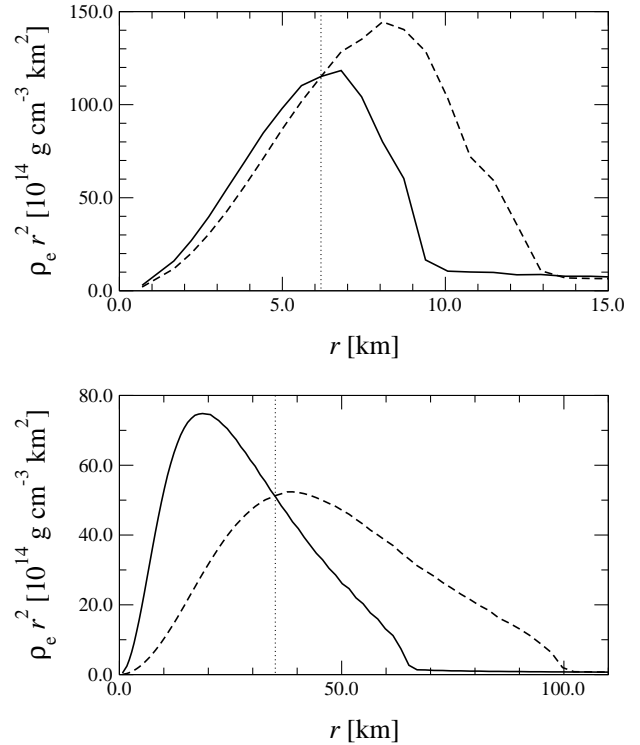}}
  \caption{Weighted radial density profile $ \rho_{\rm e} r^2 $ along
    the equator (solid lines denote relativistic models; dashed lines
    denote Newtonian models) close to the time of core bounce for model
    A3B2G4$ _{\rm soft}$ (upper panel; $ t = 39.25 {\rm\ ms} $) and
    model A2B4G1 (lower panel; $ t = 99.75 {\rm\ ms} $),
    respectively. The crossing radii are indicated by the vertical
    dotted lines.}
  \label{fig:weighted_density}
\end{figure}

The effects of the reduced density in the outer regions of the inner
core in relativistic gravity on the gravitational wave signal can be
inferred from Fig.~\ref{fig:weighted_density}, which shows the
equatorial density profile weighted by a factor $ r^2 $. Inside the
crossing radius $ r \approx 6 {\rm\ km} $, the profiles of the
weighted density $ \rho r^2 $ of model A3B2G4$ _{\rm soft}$ (upper
panel) coincide closely in both relativistic and Newtonian gravity. On
the other hand, for $ 6 {\rm\ km} \lsim r \lsim 13 {\rm\ km} $ the
weighted density $ \rho r^2 $ of the Newtonian model is significantly
larger than that of the relativistic model. Consequently, as the
density crossing occurs at all polar angles, the resulting quadrupole
moment (Eq.~\ref{eq:standard_quadrupole_formula}) is larger, too. In
order to see whether the enhanced quadrupole moment also gives rise to
a stronger gravitational wave signal, one has to consider the second
time derivative of the quadrupole moment (see
Eq.~\ref{eq:standard_quadrupole_formula}), which results from the
acceleration of the matter distribution in the core. We find that
despite the somewhat lower average acceleration in Newtonian gravity
which is reflected in the 18\% lower frequency of the gravitational
wave signal, the large quadrupole moment of the more extended density
distribution results in a maximum signal amplitude
$ |A^{\rm E2}_{20}|_{\rm max} $, which is 26\% larger than that of the
centrally denser compact remnant in relativistic gravity (see
lower panel of Fig.~\ref{fig:comparison_regular_collapse} and
Table~\ref{tab:models_summary}).

For the multiple bounce model A2B4G1, the situation is similar (lower
panel). In relativistic gravity the weighted density $ \rho r^2 $ is
enhanced in the central parts of the inner core, and reduced in the
outer parts compared to the Newtonian profile. Contrary to model
A3B2G4$ _{\rm soft} $, the relativistic density increase close to the
center has a strong effect on the weighted density profile inside the
crossing radius $ r \approx 35 {\rm\ km} $. Nevertheless, and despite
the higher accelerations (i.e., larger second time derivatives) of the
coherently oscillating inner core in relativistic gravity, the smaller
densities at larger radii again yield a smaller gravitational wave
signal (see lower panel of
Fig.~\ref{fig:comparison_multiple_bounce_collapse}). However, in
relativistic gravity the increase of both the central density (718\%
vs.\ 36\%) and the frequency (41\% vs.\ 23\%) are larger in model
A3B2G4$ _{\rm soft} $ compared to model A2B4G1. Thus, the maximum
signal amplitude $ |A^{\rm E2}_{20}|_{\rm max} $ of the former model
is only reduced by 16\% instead of 21\% in model A2B4G1 when changing
from Newtonian to relativistic gravity (see
Table~\ref{tab:models_summary}).

\begin{figure}
  \resizebox{\hsize}{!}{\includegraphics{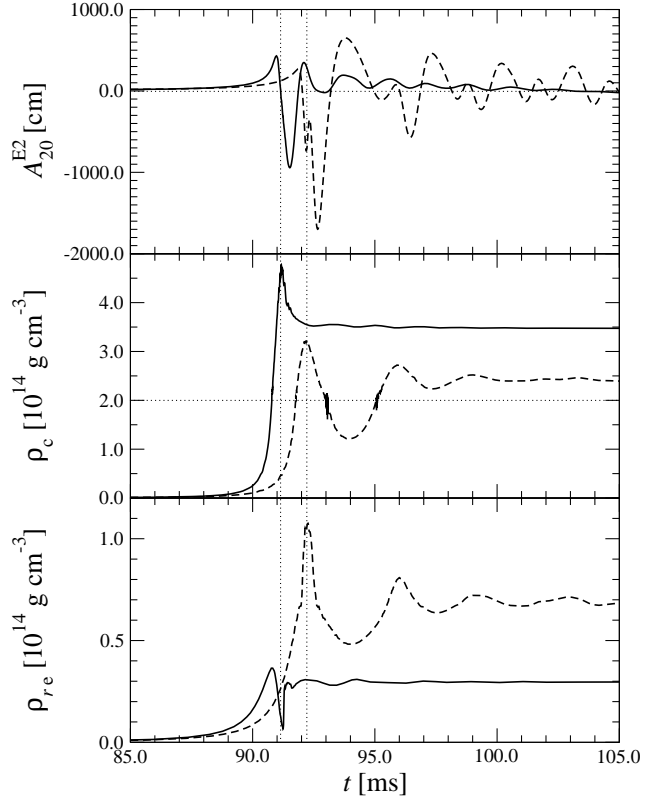}}
  \caption{Effects of a coherent motion of the inner core on the
    gravitational wave signal in the relativistic (solid lines) and
    Newtonian (dashed lines) simulation of model A1B1G1. In the upper
    panel the evolution of the gravitational wave amplitude
    $ A^{\rm E2}_{20} $ is plotted. The middle and lower panels show
    the evolution of the central density $ \rho_{\rm c} $ and of the
    equatorial density $ \rho_{r \rm\,e} $ evaluated at a radius
    $ r = 13.0 {\rm\ km} $, respectively. Contrary to the relativistic
    simulation, the inner core shows large amplitude oscillations in
    Newtonian gravity, which causes a much larger gravitational wave
    signal. The vertical dotted lines indicate the time of bounce
    $ t_{\rm b} $, and the horizontal dotted line in the middle panel
    marks nuclear matter density $ \rho_{\rm nuc} $.}
\label{fig:large_versus_small_signal}
\end{figure}

The above considerations also explain the much weaker gravitational
wave signal of model A1B1G1 in case of relativistic gravity
(see upper panel in Fig.~\ref{fig:large_versus_small_signal}). In
general relativity the central parts of the inner core collapse to
almost $ 2.5 \rho_{\rm nuc} $, and rapidly settle down to an
equilibrium state with constant density after a single bounce (middle
panel). Since the matter is more concentrated towards the center, the
density is relatively small in the outer parts of the core (lower
panel). The fluid motion in the outer regions of the core near
$ r = 13.0 {\rm\ km}$, where the shock forms, is not in phase with the
fluid motion in the inner regions (see two lower panels in
Fig.~\ref{fig:large_versus_small_signal}). In Newtonian gravity the
inner core is much less compact. It has a lower central density
(middle panel), but a much higher density at $ r = 13.0 {\rm\ km} $
(lower panel). In this case the entire inner core oscillates
coherently with large amplitude motions, which give rise to a large
time variation of the quadrupole moment and thus a large gravitational
wave signal, the maximum amplitude being almost twice as large as the
corresponding one in relativistic gravity (see
Table~\ref{tab:models_summary}).

\subsection{Change of the collapse dynamics}
\label{subsec:change_of_dynamics}

In the comprehensive study of Newtonian rotational core collapse
carried out by \citet{zwerger_97_a}, models with multiple bounces were
observed quite often. Relativistic gravity can have a qualitative
impact on the dynamics of these models. If the density increase due to
the deeper relativistic potential is sufficiently large, a collapse
which is stopped by centrifugal forces at subnuclear densities (and
thus undergoes multiple bounces) in a Newtonian simulation, becomes a
regular, single bounce collapse in relativistic gravity.

\begin{figure}
  \resizebox{\hsize}{!}{\includegraphics{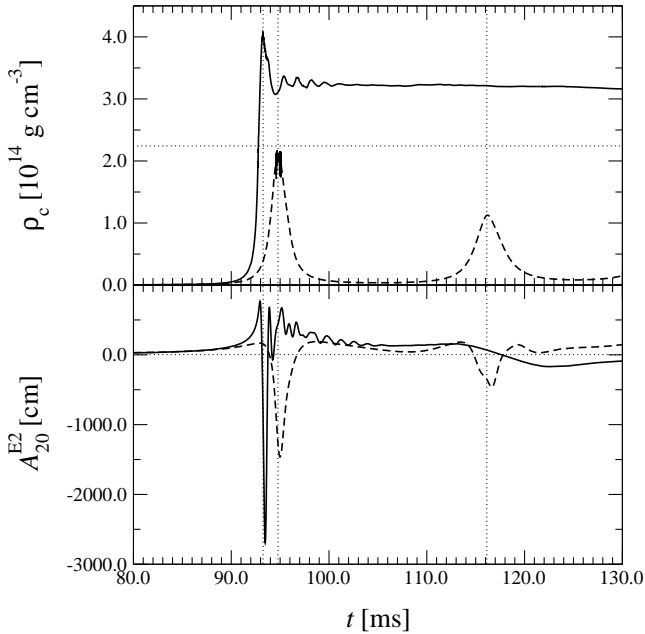}}
  \caption{Evolution of the central density $ \rho_{\rm c} $ (upper
    panel) and the gravitational wave signal amplitude
    $ A_{20}^{\rm E2} $ (lower panel) in the relativistic (solid
    lines) and Newtonian (dashed lines) simulation of model
    A1B3G1. The vertical dotted lines mark the times of peak central
    density for each bounce, and the horizontal dotted line in the
    density plot marks nuclear matter density $ \rho_{\rm nuc} $.}
  \label{fig:collapse_type_change_1}
\end{figure}

Only two of the 26 models of our sample, A2B4G1 and A3B4G2,
show unambiguous multiple bounces in relativistic gravity. These two
models share some common features. In the relativistic simulations
their peak central densities associated with the individual bounces
are much larger (up to a factor 7) compared to those of the Newtonian
runs (Table~\ref{tab:models_summary}). Consequently, the time elapsed
between subsequent bounces decreases by a factor of up to 4. These
type~II models also show the effect of density crossing (see
Section~\ref{subsec:core_compactness}). As a result, despite the
increase of the central density and the decrease of the oscillation
timescale (giving rise to a larger average acceleration of the inner
parts of the core), the maximum gravitational wave amplitude is lower
for the relativistic models.

When the central densities become very large, relativistic gravity can
even change the collapse dynamics qualitatively, i.e.\ the collapse
type is altered with respect to the Newtonian case (see
Table~\ref{tab:models_summary}). We have observed mostly
transitions from a Newtonian type~II (multiple bounce collapse) to a
relativistic type~I (regular collapse) case, and less frequently
transitions from or to type~I/II. A typical example is model A1B3G1
(Fig.~\ref{fig:collapse_type_change_1}). The change of type is clearly
visible both in the evolution of the central density (upper panel) and
of the gravitational wave signal (lower panel).

Centrifugal forces cause most of the type~II multiple bounce models of
\citet{zwerger_97_a} to bounce at densities below
$ \rho_{\rm nuc} $. Only a few models collapse to nuclear matter
density, the largest maximum density at bounce being
$ \rho_{\rm max\,b} = 3.09 \times 10^{14} {\rm\ g\ cm}^{-3} $ in the
case of model A4B1G2. Accordingly, in the two relativistic type~II
models the maximum bounce density does not exceed this
threshold. Whenever the deeper relativistic gravitational potential
drives the maximum density in the core beyond the threshold of
$ \rho_{\rm max\,b} \gsim 3 \times 10^{14} {\rm\ g\ cm}^{-3} $, a type
transition occurs (see Table~\ref{tab:models_summary}).

A collapse type transition does not only alter the signal waveform
qualitatively, but also has important consequences for the maximum
gravitational wave signal amplitude, as e.g.\ in the case of model
A1B1G1 (Fig.~\ref{fig:large_versus_small_signal}). In this Newtonian
type~I/II model the concurrence of both high densities {\em and} a
coherent motion of the inner core yields a large gravitational wave
amplitude of $ \sim 1700 {\rm\ cm} $. As the stronger relativistic
gravitational potential leads to the almost immediate formation of a
new equilibrium state, efficiently suppressing coherent large
amplitude motions of the inner core, the maximum signal amplitude is
only $ 943 {\rm\ cm} $, which is 44\% less than in the Newtonian case.

\begin{figure}
  \resizebox{\hsize}{!}{\includegraphics{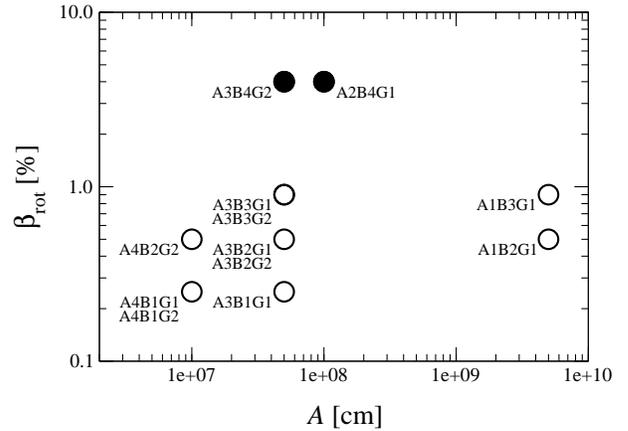}}
  \caption{Parameter space spanned by the rotation law parameter $ A $
    and the rotation rate $ \beta_{\rm rot} $ for multiple bounce
    collapse models in relativistic (normal font) and Newtonian
    (italic font) simulations. Filled circles mark models which are of
    multiple bounce type both in relativistic and Newtonian gravity,
    and empty circles mark those models which show multiple bounces in
    Newtonian gravity only.}
  \label{fig:multiple_bounce_parameters}
\end{figure}

In Newtonian gravity and for an adiabatic index $ \gamma_1 = 1.325 $
or $ \gamma_1 = 1.320 $ (i.e.\ close to the initial value
$ \gamma_{\rm ini} = 4 / 3 $) even models with rather small rotation
rates $ \beta_{\rm rot\,ini} $ and moderately differential rotation
(i.e.\ larger values of $ A $) show multiple bounces. Our simulations
demonstrate that relativistic gravity drastically narrows the region
in the parameter space of $ \beta_{\rm rot\,ini} $ and $ A $ where
multiple bounce models are
possible. Fig.~\ref{fig:multiple_bounce_parameters} shows that only
models with high rotation rates and moderate differential rotation
undergo an unambiguous type~II multiple bounce collapse.


\subsection{Rapidly and highly differentially rotating core collapse}
\label{subsec:rapid_rotation_models}

When rapidly (large $ \beta_{\rm rot\,ini} $) and strongly
differentially (small $ A $) rotating initial models collapse on a
short timescale due to a low adiabatic exponent $ \gamma_1 $, angular
momentum conservation results in a considerable spin-up of the
core. Consequently, the increasing centrifugal forces stop the
collapse before or slightly above nuclear matter density, as in
type~II multiple bounce models. However, the maximum density is
reached off-center, the topology of the density distribution in the
core being torus-like. Some initial models (e.g.\ A4B4 and A4B5)
already exhibit a toroidal structure. During collapse the toroidal
character of such a model is significantly enhanced, and the maximum
density $ \rho_{\rm max} $ can exceed the central density
$ \rho_{\rm c} $ by up to two orders of magnitude. Other rapidly and
highly differentially rotating initial models, like A3B5, A4B1 and
A4B2, in general do not change their initial oblate density
stratification during the evolution. However, for particular values of
the parameter $ \gamma_1 $, they also develop a toroidal topology
during the infall phase.

\begin{figure}
  \resizebox{\hsize}{!}{\includegraphics{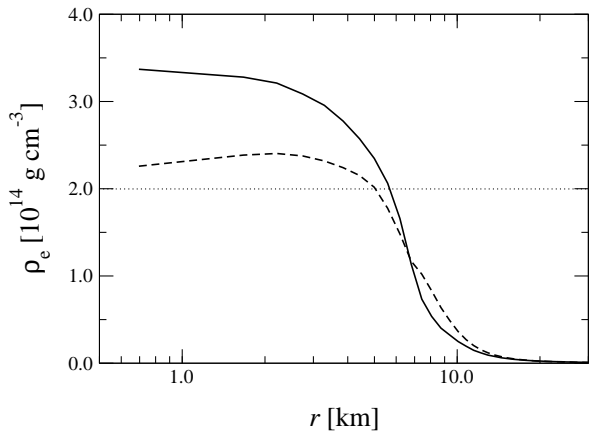}}
  \caption{Equatorial density profiles of the moderately rapidly
    rotating model A3B3G5 in case of relativistic (solid line) and
    Newtonian (dashed line) gravity, depicted after ring-down. In
    relativistic gravity the core has an oblate density structure. The
    less deep Newtonian gravitational potential leads to the
    development of a toroidal density configuration. The horizontal
    dotted line marks nuclear matter density $ \rho_{\rm nuc} $.}
  \label{fig:newtonian_only_torus}
\end{figure}

Due to their deeper gravitational potential, relativistic
configurations tend to be more compact compared to the corresponding
Newtonian ones (see Section~\ref{subsec:core_compactness}). This also
holds for relativistic models with a toroidal density stratification,
i.e.\ for models rotating rapidly and strongly differentially. These
models have off-center density maxima which exceed those of their
Newtonian counterparts (see Table~\ref{tab:models_summary}).
In some borderline models which rotate moderately fast and not too
differentially, the core develops a toroidal density stratification
only in Newtonian gravity and possesses an oblate spheroidal density
stratification in relativistic gravity. A prototype of such a case is
model A3B3G5 (see Fig.~\ref{fig:newtonian_only_torus}).

In both relativistic and Newtonian simulations of rapidly and highly
differentially rotating cores, the gravitational wave amplitudes are
generally large, because of the large quadrupole moments of toroidal
configurations (see Table~\ref{tab:models_summary}). If the rotation
rate is modest (models A4B1 and A4B2), the density in the torus at
bounce can reach high supranuclear densities in relativistic gravity.
The characteristic frequencies of the gravitational wave signal of
these toroidal models become particularly high, as they change to
regular collapse type models in relativistic gravity. The combination
of a high density torus and large accelerations (reflected by the
higher frequencies) yields very large wave amplitudes up to almost
3400~cm, which is significantly larger than in Newtonian
gravity. However, in the case of the extremely rapidly rotating models
A4B4 and A4B5, centrifugal forces surmount relativistic gravity. The
collapse type does not change, and the bounce is partially due to
centrifugal forces (type~I/II collapse). Consequently, the densities
at bounce are less extreme and the average accelerations are smaller
than in less rapidly rotating toroidal models. Therefore, the maximum
signal amplitudes $ |A^{\rm E2}_{20}|_{\rm max} $ of models A4B4 and
A4B5 are smaller than in the corresponding Newtonian models.

\begin{figure*}
  \centering \includegraphics[width=17cm]{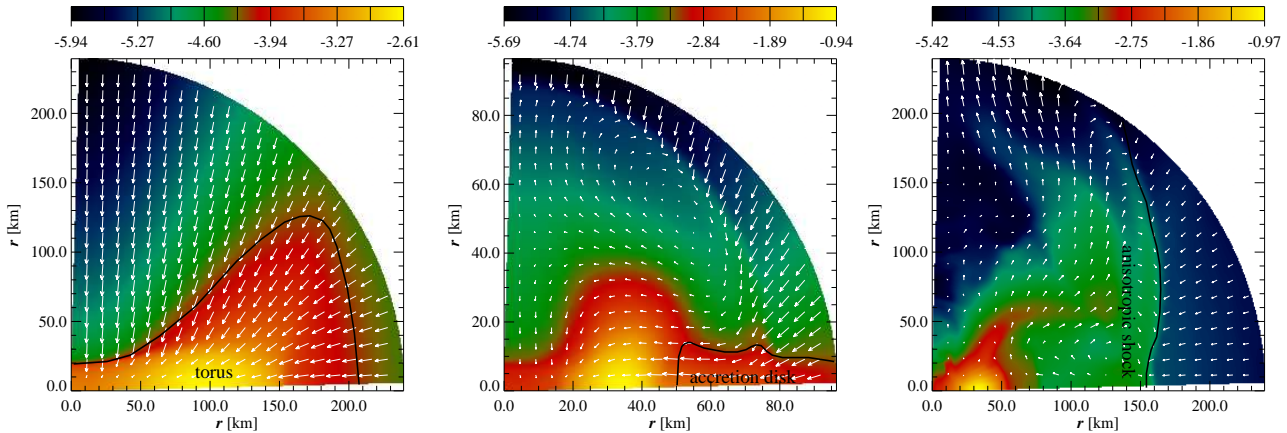}
  \caption{Formation of a torus and shock-propagation in the very
    rapidly and highly differentially rotating model A4B5G5. The three
    snapshots show color coded contour plots of the logarithm of the
    density, $ \log\,\rho_{\rm nuc} $, together with the meridional
    flow field during the infall phase at $ t = 25.0 {\rm\ ms} $ (left
    panel), shortly after the centrifugal bounce at
    $ t = 30.0 {\rm\ ms} $ (middle panel), and at
    $ t = 35.0 {\rm\ ms} $ (right panel) when the torus is surrounded
    by large scale flow vortices. The length of the velocity vectors
    is scaled to the maximum value in the plotted region. Note that
    the size of the displayed region and the color coding (which is
    given above each panel) vary from plot to plot.}
  \label{fig:shock_front_2d_torus}
\end{figure*}

Compared to models with a moderate amount of rotation, the propagation
of the shock wave is drastically different in a rapidly and
differentially rotating model, as e.g.\ in model A4B5G5
(Fig.~\ref{fig:shock_front_2d_torus}). In this model the density
maximum is located off-center already during collapse (left panel).
After core bounce the inner core settles down to an approximate
equilibrium state with a distinctive toroidal density stratification
surrounded by an equatorial disk from which matter is accreted onto
the torus (middle panel). In this accretion disk the shock encounters
relatively dense matter with large infall velocities between $ 0.1 c $
and $ 0.15 c $, and thus it propagates outward comparatively slowly
along this direction. On the other hand, the region close to the
rotation axis has been emptied of matter during the collapse, the
average density being orders of magnitude lower than close to the
equatorial plane. Therefore, here the shock propagates out very
rapidly (right panel). As a result, the flow velocity in the
meridional plane exhibits a jet-like structure, with matter being
ejected from the inner core predominantly along the rotation axis.


\subsection{Evolution of the rotation rate}
\label{subsec:rotation_rate}

\begin{figure}
  \resizebox{\hsize}{!}{\includegraphics{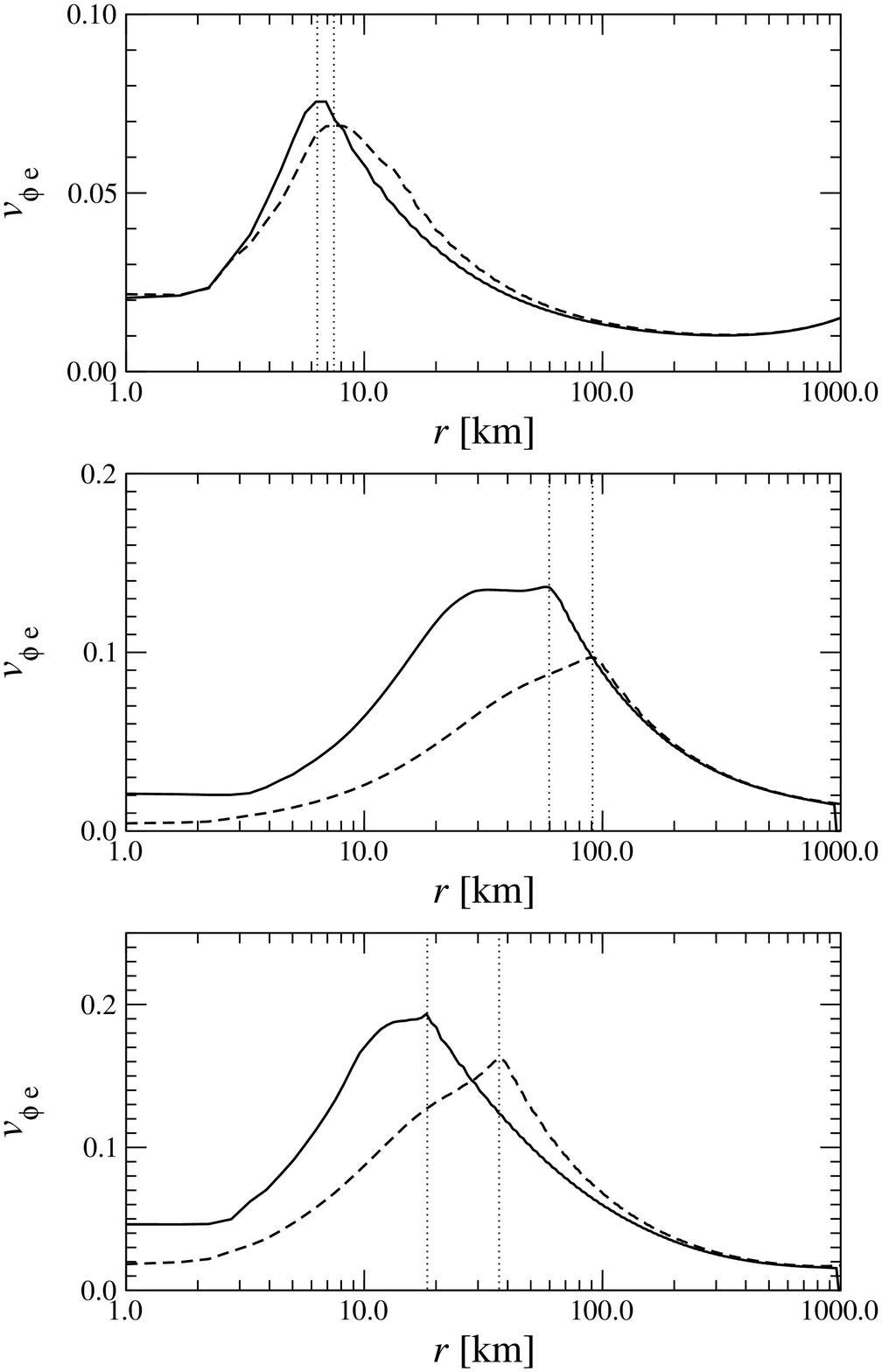}}
  \caption{Radial profiles of the rotation velocity $ v_\phi $ in the
    equatorial plane plotted shortly after core bounce obtained in the
    relativistic (solid lines) and Newtonian (dashed lines) simulation
    of models A1B3G5 (upper panel), A2B4G1 (middle panel) and A1B3G1
    (lower panel). The vertical dotted lines mark the radius of
    maximum rotation velocity.}
  \label{fig:spinup}
\end{figure}

The influence of relativistic gravity on the evolution of the rotation
rate during core collapse is demonstrated in Fig.~\ref{fig:spinup},
which shows the radial profiles of $ v_\phi = \sqrt{v_3 v^3} $ at the
time of core bounce for several representative models. Both the
maximum and average rotation velocities are higher in the relativistic
models, and the maximum of $ v_\phi $ is shifted to smaller
radii. Note that this maximum is located near the edge of the
compact remnant. The differences are most pronounced for model
A1B3G1 (lower panel), which changes from multiple bounce collapse to
regular collapse due to relativistic effects (see
Section~\ref{subsec:change_of_dynamics}). Model A1B3G1 also shows the
largest rotation velocities of all our models, $ v_\phi $ approaching
$ 0.2 c $ at $ r \approx 15 {\rm\ km} $ during core bounce. This
corresponds to a rotation period of $ T \sim 1.6 {\rm\ ms} $ at the
edge of the inner core. When collapsing from the same initial model
(A1B3), but imposing a larger initial reduction of the adiabatic index
to $ \gamma_1 = 1.280 $ (i.e.\ model A1B3G5; upper panel in
Fig.~\ref{fig:spinup}), the spin-up is less and $ v_\phi \lsim 0.08 c $.
This result can be explained as follows: In model A1B3G1 the reduced
adiabatic index $ \gamma_1 = 1.325 $ is very close to $ 4 / 3 $, i.e.\
a large fraction of the core collapses nearly homologously
\citep{goldreich_80_a}. However, this does not hold for model A1B3G5,
where the outer parts of the core fall much slower than in model
A1B3G1, and hence also fall less far in radius until the time of
bounce. Consequently, the spin-up of mass zones located initially at
the same radial position in both models is less in model A1B3G5 than
in model A1B3G1 except for the very center of the core.

Assuming angular momentum conservation for the core, the rotation rate
$ \beta_{\rm rot} $ scales like the inverse of the radius of the core
(see, e.g., \citet{shapiro_83_a}),
\begin{equation}
  \beta_{\rm rot} = \frac{E_{\rm rot}}{|E_{\rm pot}|} \sim
  \frac{J^2}{M^3 R} \propto \frac{1}{R},
  \label{eq:rotation_rate_scaling}
\end{equation}
where $ E_{\rm rot} $, $E_{\rm pot} $, $ M $, $ R $ and $ J $ are the
rotational energy, potential energy, mass, radius and angular momentum
of the core, respectively. 

Knowing the behavior of the evolution of $ \beta_{\rm rot} $ during
core collapse is important for several reasons. Firstly, the rotation
state of the inner core after bounce determines the rotation rate and
profile of the compact remnant provided no significant fall-back of
matter occurs later in the evolution. Pulsar observations show that
neutron stars can spin with very high rotation periods of the order of
milliseconds (see \citet{lorimer_01_a} and references therein).
However, such millisecond pulsars are believed to be born not as fast
rotators, but are thought to be old objects which are spun up via mass
transfer from a companion star much later in their evolution.
Observations of pulsars in young supernova remnants (see, e.g.\
\citet{kaspi_00_a}) provide only upper limits on the rotational state
of newborn neutron stars, because significant angular momentum losses
may have occurred since their formation. Presently, the fastest known
young pulsar is the 16~ms X-ray pulsar in the Crab-like supernova
remnant N157B \citep{marshall_98_a}. Secondly, at sufficiently high
rotation rates the core will become unstable against triaxial
perturbations due to various mechanisms (see \citet{stergioulas_98_a}
and references therein). Such nonaxisymmetric rotating configurations
are an additional source of gravitational radiation.

\begin{figure}
  \resizebox{\hsize}{!}{\includegraphics{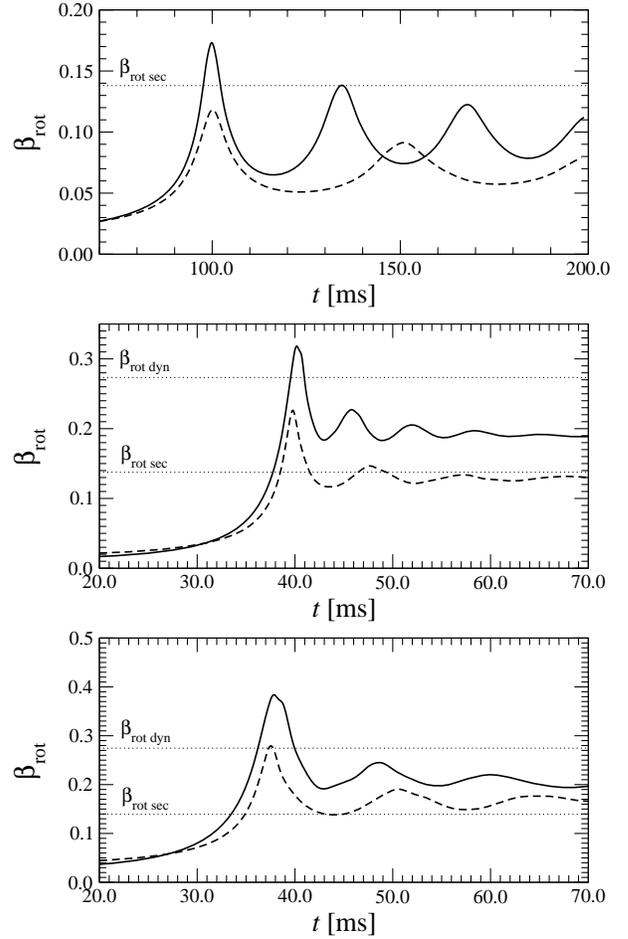}}
  \caption{Evolution of the rotation rate $ \beta_{\rm rot} $ in the
    relativistic (solid lines) and Newtonian (dashed lines)
    simulations of models A2B4G1 (upper panel), A4B4G4 (middle panel)
    and A4B5G4 (lower panel). The horizontal lines mark the critical
    rotation rates above which MacLaurin spheroids are secularly
    ($ \beta_{\rm rot\ sec} $) and dynamically
    $ \beta_{\rm rot\ dyn} $ unstable against triaxial perturbations
    in Newtonian gravity.}
  \label{fig:rotation_rate}
\end{figure}

For uniformly rotating, constant density MacLaurin spheroids in
Newtonian gravity, the threshold values for the development of
triaxial instabilities on secular and dynamic timescales are
$ \beta_{\rm rot\,sec} \approx 13.8\% $ and
$ \beta_{\rm rot\,dyn} \approx 27.4\% $, respectively (see, e.g.,
\citet{tassoul_78_a}).

The secular instability, which is driven for example by gravitational
radiation or viscosity, has been analyzed in compressible stars, both
uniformly and differentially rotating, within linear perturbation
theory in full general relativity by \citet{bonazzola_96_a,
bonazzola_98_a}, \citet{stergioulas_98_b}, and
\citet{yoshida_99_a}. The analysis shows that $ \beta_{\rm rot\,sec} $
depends on the compactness of the star, on the rotation law, and on
the dissipative mechanism. Whereas $ \beta_{\rm rot\,sec} < 0.14 $ for
the gravitational radiation-driven instability (for extremely compact
or strongly differentially rotating stars $\beta_{\rm rot\,sec} <
0.1$), secular instabilities driven by viscosity occur for larger
rotation rates, i.e.\ for values $ \beta_{\rm rot\,sec} > 0.14 $, when
the configurations become more compact.

The dynamic instability, on the other hand, is driven by hydrodynamics
and gravity. The onset of the dynamic bar-mode instability has been
investigated in full general relativity only recently
\citep{shibata_00_b, saijo_01_a}, because it requires fully nonlinear
multidimensional simulations (for Newtonian investigations see, e.g.,
\citet{new_00_a}, and references therein). The simulations show that
in relativistic gravity the critical rotation rates for the onset of
the dynamic bar-mode instability are only slightly smaller
($ \beta_{\rm rot\,dyn} \sim 0.24 - 0.25 $) than in the Newtonian
case, and depend only very weakly on the degree of differential
rotation (within the moderate range surveyed by the investigations).

What do these general considerations concerning triaxial instabilities
imply for our models? Since the rotation rates are larger in
relativistic gravity than in the Newtonian case for the same initial
conditions, we expect that the criteria for the development of dynamic
and secular triaxial instabilities are fulfilled in more models and
for longer time intervals in relativistic gravity. To see whether this
statement is correct we consider the evolution of the rotation rate
$ \beta_{\rm rot} $ for models A2B4G1, A4B4G4 and A4B5G4 (see
Fig.~\ref{fig:rotation_rate}). In all these models
$ \beta_{\rm rot} $ reaches its maximum value at the time of
bounce. If the core experiences multiple bounces, this is reflected in
multiple peaks of $ \beta_{\rm rot} $ (see upper panel). As expected,
the rotation rate in the relativistic models exceeds that of the
corresponding Newtonian ones. This holds both for the maximum value
and the average value of the rotation rate.

Using the Newtonian thresholds $ \beta_{\rm rot\,sec} $ and
$ \beta_{\rm rot\,dyn} $ (see discussion above) to judge whether a
model is prone to developing triaxial instabilities, we see that
relativistic effects do not cause an important change in the case of
model A2B4G1 (upper panel of Fig.~\ref{fig:rotation_rate}), because
$ \beta_{\rm rot} > \beta_{\rm rot\,sec} $ for less than 10~ms. On the
other hand, the behaviour of the rapidly rotating model A4B4G4 (middle
panel) is quite different. Its rotation rate exceeds
$ \beta_{\rm rot\,sec} $ near bounce time and remains above that value
after core bounce in relativistic gravity (even exceeding
$ \beta_{\rm rot\,dyn} $ for a short time), while in Newtonian gravity
it falls below $ \beta_{\rm rot\,sec} $ again after ring-down. Since
relativistic effects tend to lower the threshold value
$ \beta_{\rm rot\,sec} $ in case of the gravitational radiation-driven
instability (see discussion above), this model demonstrates that
Newtonian simulations can yield incorrect predictions about the
development of secular instabilities. On the other hand, in the
extremely rapidly and differentially rotating model A4B5G4 (lower
panel), secular instabilities are likely to develop in both Newtonian
and relativistic gravity. However, instabilities on a dynamic
timescale should only appear in the latter case, where
$ \beta_{\rm rot} > \beta_{\rm rot\,dyn} $ for almost 4~ms. We note
that for a neutron star with an average density of
$ \sim \rho_{\rm nuc} $ and a radius of $ \sim 20 {\rm\ km} $, the
dynamical timescale (=\,sound crossing timescale) is roughly 1~ms.

Note that in this discussion $ \beta_{\rm rot} $ is the ratio of
rotational energy to potential energy of the entire core. Strictly
speaking, the criteria for triaxial instabilities should only be
evaluated for the rotation rate $ \beta_{\rm rot\,ic} $ of the inner
core (which eventually becomes the compact remnant), because the
outer core is a separate entity as far as the dynamics is concerned.
However, due to the rapid drop of density and rotation velocity
observed in all models at the outer boundary of the inner core, the
outer core contributes only little to the total rotation rate. Thus,
we can safely assume that
$ \beta_{\rm rot\,ic} \approx \beta_{\rm rot} $. This behavior has
also been observed in the Newtonian core collapse simulations of
\citet{zwerger_95_a}.

Despite the knowledge of the evolution of $ \beta_{\rm rot} $ during
core collapse, predictions about the actual development of triaxial
instabilities are difficult. The thresholds originate from a linear
stability analysis and involve restrictions (e.g.\ homogeneous
density, uniform rotation, etc.). Unfortunately, we cannot simulate
the actual nonlinear growth of triaxial instabilities, because our
code is restricted to axisymmetric flows. Therefore, we have begun to
extend our code to three spatial dimensions. The results of the 3D
simulations planned to be performed with the extended code will be
presented in a future publication. Still, at present our results may
help to predict the behavior and consequences of $ \beta_{\rm rot} $
of fully three-dimensional models \citep{houser_94_a, smith_96_a,
rampp_98_a}.

For large values of $ \beta_{\rm rot} $ the neutron star is highly
nonspherical. In this regime we expect deviations of the CF metric
from the exact metric in the range of a few percent (see Section~5.5
in Paper~I). Additionally, in the equation for $ \beta_{\rm rot} $
(see Sections~1 and 5.4 in Paper~I), the potential energy
$ E_{\rm pot} $ is determined by the (small) difference between the
gravitational mass $ E_{\rm grav} $ and proper mass
$ E_{\rm proper} $. Even for strongly gravitating systems like neutron
stars, this difference is only about 10\% of the individual
terms. Thus, if the values for $ E_{\rm grav} $ and $ E_{\rm proper} $
are subject to an error of a few percent in the CFC approximation, we
expect $ \beta_{\rm rot} $ to deviate by several 10\% relative to its
value in an exact spacetime. In simulations of rapidly rotating
neutron stars in equilibrium we could observe relative deviations
between $ \beta_{\rm rot\,exact} $ and $ \beta_{\rm rot\,CFC} $ of
about 15\%. Owing to the uncertainties with the stability thresholds
and the calculation of the rotation rate in the CFC approximation, the
above predictions from the evolution of $ \beta_{\rm rot} $ for the
actual development of triaxial instabilities should be regarded as an
estimate.


\section{Gravitational wave emission}
\label{sec:gw_emission}


\subsection{Gravitational wave energy spectrum and emission}
\label{subsec:gw_energy_spectrum}

A detailed discussion of the spectra of gravitational waves emitted
during Newtonian rotational core collapse, and their dependence on the
collapse type and dynamics can be found in the work of
\citet{zwerger_97_a}. In the quadrupole approximation and in
axisymmetry, the gravitational radiation field is solely determined by
the amplitude $ A^{\rm E2}_{20} $ (see
Appendix~\ref{sec:wave_extraction}). The total energy radiated away by
gravitational waves is then given by
\begin{equation}
  E_{\rm rad\,tot} = \frac{1}{32 \pi} \int_{- \infty}^\infty \left|
  \frac{{\rm d} A^{\rm E2}_{20}}{{\rm d}t} \right|^2 {\rm d}t.
  \label{eq:total_radiated_energy_time}
\end{equation}
By replacing the gravitational wave amplitude in the time domain
$ A^{\rm E2}_{20} (t) $ by its Fourier transform in the frequency
domain $ \hat{A}^{\rm E2}_{20} (\nu) $, the total radiated energy
$ E_{\rm rad\,tot} $ can be expressed as a frequency integral:
\begin{equation}
  E_{\rm rad\,tot} = \frac{1}{16 \pi} \int_{0}^\infty \nu^2
  \left| \hat{A}^{\rm E2}_{20} \right|^2 {\rm d}\nu.
  \label{eq:total_radiated_energy_frequency}
\end{equation}
The spectral energy distribution is then
\begin{equation}
  \frac{{\rm d}E_{\rm rad}}{{\rm d}\nu} = \frac{1}{16 \pi} \nu^2 \left|
  \hat{A}^{\rm E2}_{20} \right|^2.
  \label{eq:spectral_energy}
\end{equation}

\begin{figure}
  \resizebox{\hsize}{!}{\includegraphics{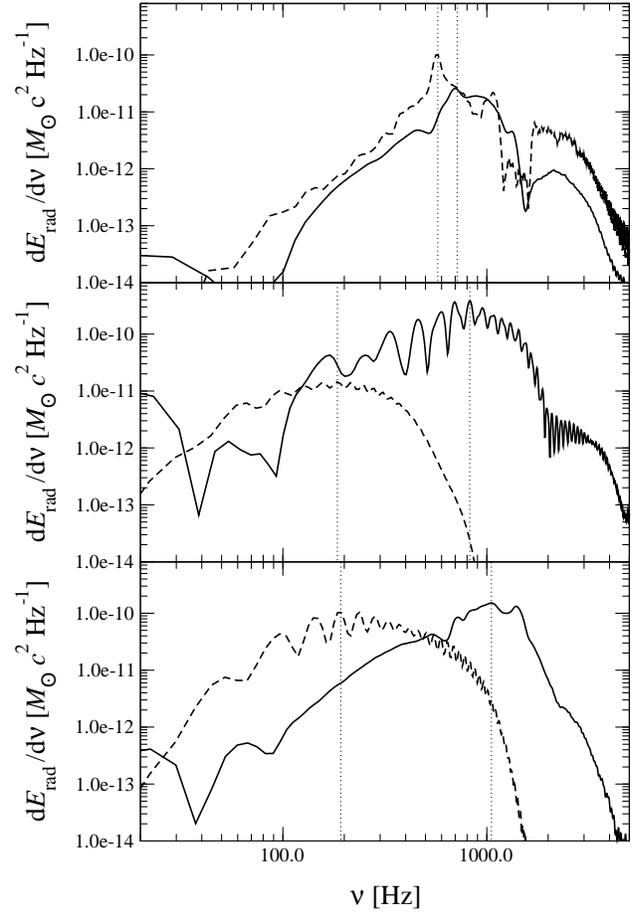}}
  \caption{Spectral energy distribution
    $ {\rm d}E_{\rm rad} / {\rm d}\nu $ of the gravitational wave
    signal in the relativistic (solid lines) and Newtonian (dashed
    lines) simulations of models A1B3G3 (upper panel), A3B3G1 (middle
    panel) and A1B3G1 (lower panel). The energy spectrum of model
    A1B3G3, with a strong low frequency part and a weaker high
    frequency part, is typical for the signal of a regular
    collapse. Model A3B3G1 is a multiple bounce collapse model in both
    Newtonian and relativistic gravity. In its energy spectrum a broad
    low frequency peak dominates. In model A1B3G1 relativistic effects
    change the collapse type from multiple bounce collapse to regular
    collapse. The energy spectrum reflects this change. The left
    (right) vertical dotted lines mark the maxima of the energy
    spectrum in Newtonian (relativistic) gravity. Due to higher
    average central densities in the relativistic simulations, the
    maximum of the energy spectrum is shifted to higher frequencies.}
  \label{fig:energy_spectrum}
\end{figure}

As a characteristic spectral measure of the gravitational wave signal
we define the frequency $ \nu_{\rm max} $ at which the energy spectrum
$ {\rm d}E_{\rm rad} / {\rm d}\nu $ has its maximum (see also
\citet{zwerger_97_a}).

As discussed in Section~\ref{subsec:core_compactness}, the average and
peak central densities of our relativistic models are larger than
those of the Newtonian ones. This leads to an increase of the
oscillation frequencies of the inner core during the ring-down phase
in a regular collapse situation, and to a decrease of the time
interval between consecutive bounces in multiple bounce models.
Therefore, the gravitational wave spectrum is shifted towards higher
frequencies in the relativistic models. This shift can be as large as
a factor 5 in frequency (see Table~\ref{tab:models_summary} and lower
panel of Fig.~\ref{fig:energy_spectrum}). The models with the largest
frequency shift are those where the deeper relativistic gravitational
potential changes a multiple bounce collapse into a regular collapse
(see Section~\ref{subsec:change_of_dynamics}). We find that in
Newtonian gravity only three models have a higher characteristic
frequency $ \nu_{\rm max} $ than the corresponding relativistic ones.
However, the energy spectra of these models show many local maxima
near $ \nu_{\rm max} $ having almost the same spectral energy
density. Fitting the energy spectra of the three models by a smooth
function yields again a higher value of $ \nu_{\rm max} $ in case of
relativistic gravity.

The energy spectra of three models are plotted in
Fig.~\ref{fig:energy_spectrum}. They are obtained by Fourier
transforming the gravitational signal. A Welch filter is applied in
the Fourier transformation to reduce the noise caused by the finite
length of the wave signal \citep{press_92_a}. To compensate for the
power loss due to filtering, the spectrum is rescaled by the
requirement that the total energy in the time domain,
Eq.~(\ref{eq:total_radiated_energy_time}) equals that of the frequency
domain, Eq.~(\ref{eq:total_radiated_energy_frequency}). The energy
spectra are similar for both relativistic and Newtonian simulations in
cases where the collapse type does not change (see upper and middle
panel in Fig.~\ref{fig:energy_spectrum}). On the other hand, when a
Newtonian model exhibiting multiple bounces changes to a regular
collapse model in relativistic gravity, like e.g.\ model A3B3G1, the
shape of its energy spectrum is shifted significantly (see lower panel
in Fig.~\ref{fig:energy_spectrum}).

By integrating the spectral energy distribution,
Eq.~(\ref{eq:spectral_energy}), over all frequencies, one obtains the
total energy radiated away by the system in gravitational waves, as
given in Eq.~(\ref{eq:total_radiated_energy_frequency}). However, when
analyzing numerical data this method is inappropriate, because high
frequency noise contributes significantly to $ E_{\rm rad\,tot} $. For
most models it is impossible to unambiguously define a cutoff
frequency beyond which there exists only noise, as $ E_{\rm rad} $
does not drop sharply at some specific frequency. On the other hand,
applying a filter (like the Welch filter used in
Fig.~\ref{fig:energy_spectrum}) causes a leakage of spectral energy
density, and thus leads to a underestimate of $ E_{\rm rad\,tot} $.
Therefore, we use Eq.~(\ref{eq:total_radiated_energy_time}) to
calculate $ E_{\rm rad\,tot} $. 

We find that the total energy (measured in units of
$ M_{\odot} c^2 $) radiated in form of gravitational waves lies in the
range
$ 3.0 \times 10^{-10} \le E_{\rm rad\,tot} \le 3.7 \times 10^{-7} $
for the relativistic models, and in the range
$ 3.8 \times 10^{-10} \le E_{\rm rad\,tot} \le 1.5 \times 10^{-7} $
for the Newtonian ones. Averaged over all models,
$ E_{\rm rad\,tot} = 8.2 \times 10^{-8} $ in the relativistic case,
and $ E_{\rm rad\,tot} = 3.6 \times 10^{-8} $ in the Newtonian one,
which corresponds to an increase of 128\% due to relativistic effects.


\subsection{Gravitational wave amplitudes and frequencies}
\label{subsec:gw_amplitudes_and_frequencies}

\begin{figure*}
  \sidecaption
  \includegraphics[width=12cm]{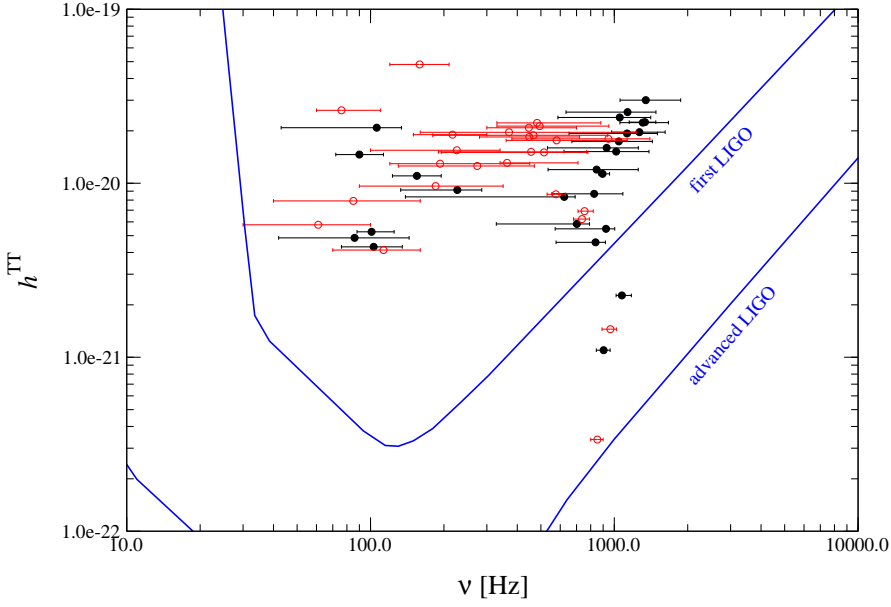}
  \caption{~Prospects of detection of the gravitational wave signal
    from axisymmetric rotational supernova core collapse in relativistic
    (black filled circles) and Newtonian (red unfilled circles)
    gravity. The figure gives the (dimensionless) gravitational wave
    amplitude $ h^{\rm TT} $ and the frequency range for all 26
    models. For a source at a distance of 10~kpc the signals of all
    models are above the burst sensitivity of the LIGO~I detector
    (except for some low amplitude, high frequency models), and well
    above that of the LIGO~II interferometer. The burst sensitivity
    gives the r.m.s.\ noise amplitude
    $ h_{\rm rms} \sim \sqrt{\nu S(\nu)}$ over a bandwith of width
    $\nu$ at a frequency $ \nu $ for the instrument noise power
    spectral density $ S(\nu) $. The error bars mark the frequency
    range inside which the spectral energy density is within 50\% of
    its peak value.}
  \label{fig:sensitivity_all_models}
\end{figure*}

\begin{figure*}
  \sidecaption
  \includegraphics[width=12cm]{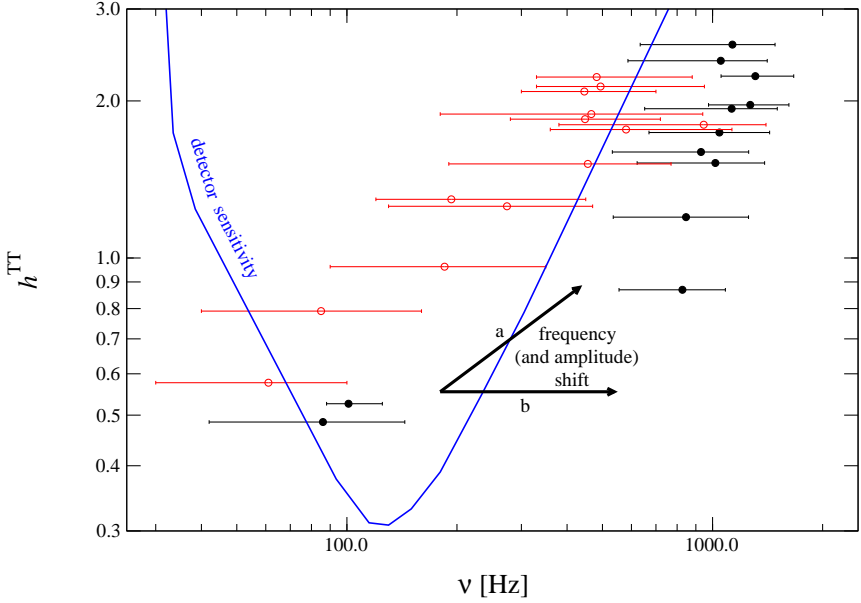}
  \caption{~Impact of the frequency and wave amplitude shift on the
    signal detectability for relativistic models which experience
    multiple bounces in Newtonian gravity. In the
    ``best case'', both amplitude and frequency increase in relativistic
    gravity (a). However, on average only the frequency increases
    (b). In neither case the shift is as steep as the high frequency
    sensitivity curve of an interferometer detector, and thus a
    relativistic model may fall out of the sensitivity window of the
    detector, if the Newtonian model is just within detection
    range. Note that the vertical scale is arbitrary. Symbols and
    error bars as in Fig.~\ref{fig:sensitivity_all_models}.}
  \label{fig:sensitivity_multiple_bounce_models}
\end{figure*}

The distribution of the dimensionless signal amplitude seen by an
observer located in the equatorial plane of the source, i.e.\ the
gravitational wave strain (see Eq.~(\ref{eq:h_tt}) with
$ \theta = \pi / 2 $)
\begin{equationarray}
  h^{\rm TT} & = & \frac{1}{8} \sqrt{\frac{15}{\pi}}
  \frac{A_{20}^{\rm E2}}{r} \nonumber \\
  & = & 8.8524 \times 10^{- 21} 
  \left( \frac{A_{20}^{\rm E2}}{10^3 {\rm\ cm}} \right)
  \left( \frac{10 {\rm\ kpc}}{r} \right),
  \label{eq:conversion_a_quad_to_h_tt}
\end{equationarray}%
is shown in Fig.~\ref{fig:sensitivity_all_models} for all 26 models,
scaled to a distance of $ r = 10 {\rm\ kpc} $ from the source. The
corresponding wave amplitudes $ A^{\rm E2}_{20} $ are listed in
Table~\ref{tab:models_summary}.

A comparison of Fig.~\ref{fig:sensitivity_all_models} with Fig.~8 in
\citet{zwerger_97_a} proves that our sample, encompassing 25 of the 78
models considered in the parameter study of \citet{zwerger_97_a} -- we
have simulated the additional model A3B2G4$ _{\rm soft} $ -- covers
almost the same range of wave amplitudes
($4 \times 10^{-21} \le h^{\rm TT} \le 3 \times 10^{-20} $) and
frequencies ($ 60 {\rm\ Hz} \le \nu \le 1000 {\rm\ Hz} $). Thus,
relativistic effects have no significant impact on the gross
distribution of expected signals in the amplitude--frequency
diagram. Only in the high frequency, low amplitude region of the
diagram, i.e.\ for models which collapse rapidly and have type~I or
III waveforms, the weaker signal amplitudes cause a discernible change
in the model distribution. The bulk of the models falls into the range
of detectability of the LIGO~I interferometer \citep{gustafson_99_a}
provided the source is located at a distance of less than
$ 10 {\rm\ kpc} $ (see Fig.~\ref{fig:sensitivity_all_models}).

For all models which are of the same type in both Newtonian and
relativistic gravity, the gravitational wave signal has a lower
amplitude. When selecting those 13 models which are of type~II in
Newtonian gravity (see Section~\ref{subsec:multiple_bounce_collapse}),
the influence of relativistic effects on the signal amplitude is not
as clearcut. The two relativistic models which remain of type~II (see
Table~\ref{tab:models_summary}) yield smaller maximum amplitudes
$ |A^{\rm E2}_{20}|_{\rm max} $. If the collapse type changes due to the
higher central densities reached in relativistic gravity, some models
exhibit significantly higher signal amplitudes than in Newtonian
gravity (e.g.\ $ +84\% $ for model A1B3G1), whereas others possess
smaller signal amplitudes (e.g.\ $ -22\% $ for model A3B1G1). In
models which change their collapse type, the increase or decrease of
the signal strength, and thus the change of location in the
amplitude--frequency diagram, depends very strongly on the rotational
state of the initial model and on the value of $ \gamma_1 $. However,
as in all other models, relativistic gravity always causes a rise of
the characteristic signal frequencies. 

The overall behaviour of the 13 models which are of type~II in
Newtonian gravity, can be summarized by considering the arithmetic
mean of their peak amplitudes and peak frequencies:
\begin{equation}
  \begin{array}{l}
    \left.
    \renewcommand{\arraystretch}{1.5}
      \begin{array}{l}
        \nu_{\rm mean}^{\rm N} = 394 {\rm\ Hz} \\
        \nu_{\rm mean}^{\rm R} = 930 {\rm\ Hz}
      \end{array}
    \right\} +136\%, \\ \\
    \left.
    \renewcommand{\arraystretch}{1.5}
      \begin{array}{l}
        h_{\rm mean}^{\rm TT~N} = 1.55 \times 10^{- 20} \\
        h_{\rm mean}^{\rm TT~R} = 1.69 \times 10^{- 20}
      \end{array}
    \right\} +9\%.
  \end{array}
  \label{eq:multiple_bounce_migration}
\end{equation}
Our simulations show that while the average gravitational wave signal
amplitude $ h^{\rm TT} $ does not change significantly, the frequency
more than doubles.

This large frequency shift can have important consequences for the
prospects of detection of gravitational waves. If the signal of a
particular model is close to the detection threshold of a detector, as
shown in Fig.~\ref{fig:sensitivity_multiple_bounce_models} for the 13
models which experience multiple bounces in Newtonian gravity, the
influence of relativity can be twofold. In the ``best case scenario''
the signal is shifted to both higher frequencies and amplitudes
(case~a). However, more frequently the location of the model will only
shift towards higher frequencies in the amplitude-frequency diagram
(case~b -- see Eq.~(\ref{eq:multiple_bounce_migration})), and
therefore in a borderline case the signal may actually leave the
sensitivity window of the detector.


\section{Summary and conclusions}
\label{sec:conclusions}

We have presented hydrodynamic simulations of relativistic rotational
supernova core collapse in axisymmetry and have analyzed the
gravitational radiation emitted by such an event. We have simulated
the evolution of 26 models in both Newtonian and relativistic
gravity. Collapse of the initial configurations, which are
differentially rotating relativistic $ 4 / 3 $-polytropes in
equilibrium, is induced by decreasing the adiabatic index to some
prescribed fixed value $ \gamma_1 $ with
$ 1.28 \le \gamma_1 \le 1.325 $. The stiffening of the equation of
state at nuclear matter density and the thermal pressure in the matter
heated by the prompt shock are simulated by means of a simplified
equation of state consisting of a polytropic and a thermal
part. Microphysics like electron captures and neutrino transport is
neglected.

Our simulations show that the three different types of rotational
supernova core collapse and gravitational waveforms identified in
previous Newtonian simulations (regular collapse, multiple bounce
collapse, and rapid collapse) are also present in relativistic
gravity. We find, however, that rotational core collapse with multiple
bounces, which occurs when the collapse is only or predominantly stopped
by centrifugal forces, is possible only in a much narrower parameter
range in relativistic simulations. Only two of the 26 models of our
sample show unambiguous multiple bounces in relativistic gravity.
The peak central densities associated with the individual bounces in
such models are much larger (up to a factor of 8) in the relativistic
models compared to the corresponding Newtonian ones. Consequently, the
time elapsed between subsequent bounces decreases by a factor of up to
4.

We further find that relativistic gravity can have a qualitative
impact on the dynamics. If the density increase due to the deeper
relativistic potential is sufficiently large, a collapse which is
stopped by centrifugal forces at subnuclear densities (and thus
undergoes multiple bounces) in a Newtonian simulation, becomes a
regular, single bounce collapse in relativistic gravity. A collapse
type transition also has important consequences for the maximum
gravitational wave signal amplitude.

For all relativistic models the peak density at bounce is larger than
in the corresponding Newtonian models, the relative increase reaching
$\sim 700\%$ in some cases. This also holds for the maximum density
of the compact remnant after ring-down in case of instant formation
of a stable equilibrium state (types~I and III). Nevertheless, we
find that only six models have also larger maximum signal amplitudes
than their Newtonian counterparts. The maximum signal amplitudes of
all other relativistic models are (up to 57\%) smaller than those of
the corresponding Newtonian ones. The reduced maximum signal strength
can be explained by the fact that the quadrupole amplitude is
determined by the {\em bulk} motion of the core rather than just by
the motion of the densest mass shells. Therefore, a core which is
more condensed in the center can give rise to a {\em smaller}
gravitational wave signal than a core which is less centrally
condensed, but which is denser and moves faster in its outer regions.
However, the fact that all relativistic models show an increased
central and reduced peripheral density of the collapsed core
relative to their Newtonian counterparts does not necessarily imply a
weaker gravitational wave signal, as in 6 of 26 relativistic models
the maximum gravitational wave amplitudes are larger than in the
corresponding Newtonian models. In all these six cases the collapse
changes from type~II (or type~I/II) to type~I when changing from
Newtonian to relativistic gravity.

Overall, the relativistic models cover almost the same range of
gravitational wave amplitudes
($ 4 \times 10^{-21} \le h^{\rm TT} \le 3 \times 10^{-20} $) and
frequencies ($ 60 {\rm\ Hz} \le \nu \le 1000 {\rm\ Hz} $) as the
corresponding Newtonian ones. For all models which are of the same
collapse type in both Newtonian and relativistic gravity, the
gravitational wave signal is of lower amplitude. If the collapse type
changes, either weaker or stronger signals are found in the
relativistic case. Averaged over all models, the total energy radiated
in form of gravitational waves is
$ 8.2 \times 10^{-8}\, M_{\odot} c^2 $ in the relativistic case, and
$ 3.6 \times 10^{-8}\, M_{\odot} c^2 $ in the Newtonian one.

For a given model, relativistic gravity can cause a large increase of
the characteristic signal frequency of up to a factor of five. This
frequency shift is an effect common to all models, and can have
important consequences for the prospects of detection of gravitational
waves. If the signal of a particular model is close to the
high-frequency detection threshold of a detector, the influence of
relativity can be twofold. In the ``best case scenario'' both the
signal frequency and amplitude increase; however, more often only the
signal frequency increases. In either case, in a borderline situation the
signal may actually leave the sensitivity window of the detector, as
the high-frequency detection threshold of a laser interferometer rises
steeply with increasing frequency. Hence, within the limits of our
physical model of rotational core collapse, relativistic effects on
average tend to decrease the prospects of detectability of
gravitational radiation by interferometric detectors. Nevertheless,
the gravitational wave signals obtained in our study are within the
sensitivity range of the first generation laser interferometer
detectors if the source is located within the Local Group. 

In our study we have restricted ourselves to simplified supernova core
models. The complicated microphysics (electron captures and other
weak interactions, equation of state) and neutrino transport processes
have all been absorbed in one model parameter ($ \gamma_1 $). Although
this is an extreme simplification, we nevertheless think that it does
not change the {\em qualitative} nature of our results. In particular,
we do not expect that more realistic (axisymmetric) core collapse
models will produce significantly stronger gravitational wave signals.
Larger initial rotation rates and lower values of the (effective)
adiabatic index $\gamma_1$, which could lead to stronger gravitational
wave signals, are most likely excluded. This is also supported by the
fact that the signal strengths of the microscopically more realistic,
but Newtonian models of \citet{moenchmeyer_91_a} agree well with those
of our simplified models. To check whether these expectations are
indeed correct we plan to incorporate a realistic equation of state
and a simplified treatment of neutrino transport into our relativistic
code in the near future.

In several models the rotation rate of the compact remnant exceeds
the critical value where MacLaurin spheroids become secularly, and in
some cases even dynamically unstable against triaxial perturbations
($ \beta_{\rm rot\,sec} \approx 13.8\% $ and
$ \beta_{\rm rot\,dyn} \approx 27.4\% $). It has been have pointed out
that if such instabilities indeed occur, a much stronger gravitational
wave signal might be produced than by cores which remain axisymmetric
(for a review, see e.g.\ \citet{thorne_95_a}). According to this idea,
the core will be transformed into a bar-like configuration that spins
end-over-end like an American football. One has further speculated,
whether the core might even break up into two or more massive pieces,
if $ \beta_{\rm rot} > \beta_{\rm rot\,dyn} $. In that case the
resulting gravitational radiation {\em could} be almost as strong as
that from coalescing neutron star binaries \citep{thorne_95_a}. The
strength of the gravitational signal sensitively depends on what
fraction of the angular momentum of the non-axisymmetric core goes
into gravitational waves, and what fraction into hydrodynamic
waves. These sound and shock waves are produced as the bar or lumps,
acting like a twirling-stick, plow through the surrounding matter. In
order to investigate these effects, three-dimensional relativistic
simulations of rotational core collapse have to be performed, which
can follow the growth of non-axisymmetric instabilities (bar modes and
gravitational wave driven unstable inertial modes). To this end
we are currently extending our code to three spatial dimensions.


\begin{acknowledgements}
  This work was finished during a visit of H.D.\ at the Universidad de
  Valencia, Spain. He would like to thank for the hospitality at the
  Departamento de Astronom\'{\i}a y Astrof\'{\i}sica. J.A.F.\
  acknowledges support from a Marie Curie fellowship by the European
  Union (MCFI-2001-00032). All computations were performed on the NEC
  SX-5/3C supercomputer at the Rechenzentrum Garching.
\end{acknowledgements}

\bibliographystyle{aa}
\bibliography{aa_paper}

\begin{thebibliography}{64}
\expandafter\ifx\csname natexlab\endcsname\relax\def\natexlab#1{#1}\fi

\bibitem[{Alcubierre {et~al.}(2000)Alcubierre, Br\"ugmann, Dramlitsch, Font,
  Papadopoulos, Seidel, Stergioulas, \& Takahashi}]{alcubierre_00_a}
Alcubierre, M., Br\"ugmann, B., Dramlitsch, T., {et~al.} 2000, Phys.\ Rev.~D,
  62, 044034

\bibitem[{Aloy {et~al.}(2000)Aloy, M\"uller, Ib\'a\~nez, Mart\'{\i}, \&
  MacFadyen}]{aloy_00_a}
Aloy, M.~A., M\"uller, E., Ib\'a\~nez, J.~M., Mart\'{\i}, J.~M., \& MacFadyen,
  A. 2000, Astrophys.~J.\ Lett., 531, L119

\bibitem[{Banyuls {et~al.}(1997)Banyuls, Font, Ib\'a\~nez, Mart\'{\i}, \&
  Miralles}]{banyuls_97_a}
Banyuls, F., Font, J.~A., Ib\'a\~nez, J.~M., Mart\'{\i}, J.~M., \& Miralles,
  J.~A. 1997, Astrophys.~J., 476, 221

\bibitem[{Blanchet {et~al.}(1990)Blanchet, Damour, \&
  Sch\"afer}]{blanchet_90_a}
Blanchet, L., Damour, T., \& Sch\"afer, G. 1990, Mon.\ Not.\ R.~Astron.\ Soc.,
  242, 289

\bibitem[{Bodenheimer \& Woosley(1983)}]{bodenheimer_83_a}
Bodenheimer, P. \& Woosley, S.~E. 1983, Astrophys.~J., 269, 281

\bibitem[{Bonazzola {et~al.}(1996)Bonazzola, Frieben, \&
  Gourgoulhon}]{bonazzola_96_a}
Bonazzola, S., Frieben, J., \& Gourgoulhon, E. 1996, Astrophys.~J., 460, 379

\bibitem[{Bonazzola {et~al.}(1998)Bonazzola, Frieben, \&
  Gourgoulhon}]{bonazzola_98_a}
---. 1998, Astron.\ Astrophys., 331, 280

\bibitem[{Bonazzola \& Marck(1993)}]{bonazzola_93_a}
Bonazzola, S. \& Marck, J.~A. 1993, Astron.\ Astrophys., 267, 623

\bibitem[{Brown(2001)}]{brown_01_a}
Brown, J.~D. 2001, in AIP Conference Proc., Vol.~575, ``Astrophysical sources
  for ground-based gravitational wave detectors'', ed. J.~M. Centrella (New
  York, USA: American Institute of Physics, Melville), 234--245

\bibitem[{Dimmelmeier {et~al.}(2001)Dimmelmeier, Font, \&
  M\"uller}]{dimmelmeier_01_a}
Dimmelmeier, H., Font, J.~A., \& M\"uller, E. 2001, Astrophys.~J.\ Lett., 560,
  L163

\bibitem[{Dimmelmeier {et~al.}(2002)Dimmelmeier, Font, \&
  M\"uller}]{dimmelmeier_02_a}
---. 2002, Astron.\ Astrophys., in press (paper~I)

\bibitem[{Evans(1986)}]{evans_86_a}
Evans, C.~R. 1986, in Dynamical spacetimes and numerical relativity, ed. J.~M.
  Centrella (Cambridge, U.~K.: Cambridge University Press), 3--39

\bibitem[{Finn(1989)}]{finn_89_a}
Finn, L.~S. 1989, in Frontiers in numerical relativity, ed. C.~R. Evans, S.~L.
  Finn, \& D.~W. Hobill (Cambridge, U.~K.: Cambridge University Press),
  126--145

\bibitem[{Finn \& Evans(1990)}]{finn_90_a}
Finn, L.~S. \& Evans, C.~R. 1990, Astrophys.~J., 351, 588

\bibitem[{Fryer \& Heger(2000)}]{fryer_00_a}
Fryer, C. \& Heger, A. 2000, Astrophys.~J., 541, 1033

\bibitem[{Fryer {et~al.}(2002)Fryer, Holz, \& Heger}]{fryer_02_a}
Fryer, C., Holz, D.~E., \& Heger, A. 2002, Astrophys.~J., 565, 430

\bibitem[{Goldreich \& Weber(1980)}]{goldreich_80_a}
Goldreich, P. \& Weber, S.~V. 1980, Astrophys.~J., 238, 991

\bibitem[{Gustafson {et~al.}(1999)Gustafson, Shoemaker, Strain, \&
  Weiss}]{gustafson_99_a}
Gustafson, E., Shoemaker, D., Strain, K., \& Weiss, R. 1999, LSC white paper on
  detector research and development, Tech. Rep. LIGO T990080-00-D

\bibitem[{Hayashi {et~al.}(1999)Hayashi, Eriguchi, \& Hashimoto}]{hayashi_99_a}
Hayashi, A., Eriguchi, Y., \& Hashimoto, M. 1999, Astrophys.~J., 521, 376

\bibitem[{Heger {et~al.}(2000)Heger, Langer, \& Woosley}]{heger_00_a}
Heger, A., Langer, N., \& Woosley, S.~E. 2000, Astrophys.~J., 528, 368

\bibitem[{Houser {et~al.}(1994)Houser, Centrella, \& Smith}]{houser_94_a}
Houser, J.~L., Centrella, J.~M., \& Smith, S.~C. 1994, Phys.\ Rev.\ Lett., 72,
  1314

\bibitem[{Imshennik \& Nadezhin(1992)}]{imshennik_92_a}
Imshennik, V.~S. \& Nadezhin, D.~K. 1992, Sov. Astron. Lett., 18, 79

\bibitem[{Janka \& M\"onchmeyer(1989)}]{janka_89_a}
Janka, H.-T. \& M\"onchmeyer, R. 1989, Astron.\ Astrophys., 226, 69

\bibitem[{Janka {et~al.}(1993)Janka, Zwerger, \& M\"onchmeyer}]{janka_93_a}
Janka, H.-T., Zwerger, T., \& M\"onchmeyer, R. 1993, Astron.\ Astrophys., 268,
  360

\bibitem[{Kaspi(2000)}]{kaspi_00_a}
Kaspi, V.~M. 2000, in ASP Conference Series, Vol.~202, "Pulsar Astronomy - 2000
  and Beyond", ed. M.~Kramer, N.~Wex, \& N.~Wielebinski (San Franciso, CA, USA:
  Astronomical Society of the Pacific), 485--490

\bibitem[{Komatsu {et~al.}(1989{\natexlab{a}})Komatsu, Eriguchi, \&
  Hachisu}]{komatsu_89_a}
Komatsu, H., Eriguchi, Y., \& Hachisu, I. 1989{\natexlab{a}}, Mon.\ Not.\
  R.~Astron.\ Soc., 237, 355

\bibitem[{Komatsu {et~al.}(1989{\natexlab{b}})Komatsu, Eriguchi, \&
  Hachisu}]{komatsu_89_b}
---. 1989{\natexlab{b}}, Mon.\ Not.\ R.~Astron.\ Soc., 239, 153

\bibitem[{Lorimer(2001)}]{lorimer_01_a}
Lorimer, D.~R. 2001, Living Rev.\ Relativity, 4, cited on 10 Dec 2001,
  http://www.livingreviews.org/\hspace{0 cm}Articles/\hspace{0
  cm}Volume4/\hspace{0 cm}2001-5lorimer/

\bibitem[{MacFadyen {et~al.}(2001)MacFadyen, Woosley, \&
  Heger}]{mac_fadyen_01_a}
MacFadyen, A.~I., Woosley, S.~E., \& Heger, A. 2001, Astrophys.~J., 550, 410

\bibitem[{Marshall {et~al.}(1998)Marshall, Gotthelf, Zhang, Middleditch, \&
  Wang}]{marshall_98_a}
Marshall, F., Gotthelf, E., Zhang, W., Middleditch, J., \& Wang, Q. 1998,
  Astrophys.~J.\ Lett., 499, L179

\bibitem[{M\"onchmeyer \& M\"uller(1989)}]{moenchmeyer_89_a}
M\"onchmeyer, R. \& M\"uller, E. 1989, in NATO ASI ``Timing neutron stars'',
  ed. H.~\"Ogelman \& E.~P.~J. van~den Heuvel (Dordrecht, Netherlands: Kluwer),
  549--572

\bibitem[{M\"onchmeyer {et~al.}(1991)M\"onchmeyer, Sch\"afer, M\"uller, \&
  Kates}]{moenchmeyer_91_a}
M\"onchmeyer, R., Sch\"afer, G., M\"uller, E., \& Kates, R.~E. 1991, Astron.\
  Astrophys., 246, 417

\bibitem[{M\"uller(1982)}]{mueller_82_a}
M\"uller, E. 1982, Astron.\ Astrophys., 114, 53

\bibitem[{M\"uller(1998)}]{mueller_98_a}
M\"uller, E. 1998, in Computational methods for astrophysical fluid flow.
  Saas-Fee Advanced Course 27, ed. O.~Steiner \& A.~Gautschy (Berlin, Germany:
  Springer), 343--494

\bibitem[{M\"uller \& Hillebrandt(1981)}]{mueller_81_a}
M\"uller, E. \& Hillebrandt, W. 1981, Astron.\ Astrophys., 103, 358

\bibitem[{M\"uller \& Janka(1997)}]{mueller_97_a}
M\"uller, E. \& Janka, H.-T. 1997, Astron.\ Astrophys., 317, 140

\bibitem[{Nakamura(1981)}]{nakamura_81_a}
Nakamura, T. 1981, Prog.\ Theor.\ Phys., 65, 1876

\bibitem[{Nakamura(1983)}]{nakamura_83_a}
---. 1983, Prog.\ Theor.\ Phys., 70, 1144

\bibitem[{Nakamura {et~al.}(1987)Nakamura, Oohara, \& Kojima}]{nakamura_87_a}
Nakamura, T., Oohara, K., \& Kojima, Y. 1987, Prog.\ Theor.\ Phys., 90, 1

\bibitem[{New {et~al.}(2000)New, Centrella, \& Tohline}]{new_00_a}
New, K.~C.~B., Centrella, J.~M., \& Tohline, J.~E. 2000, Phys.\ Rev.~D, 62,
  064019

\bibitem[{Pradier {et~al.}(2001)Pradier, Arnaud, Bizouard, Cavalier, Davier, \&
  Hello}]{pradier_01_a}
Pradier, T., Arnaud, N., Bizouard, M.-A., {et~al.} 2001, Phys.\ Rev.~D, 63,
  042002

\bibitem[{Press {et~al.}(1992)Press, Teukolsky, Vetterling, \&
  Flannery}]{press_92_a}
Press, W.~H., Teukolsky, S.~A., Vetterling, W.~T., \& Flannery, B.~P. 1992,
  Numerical recipes in C: The art of scientific programming (Cambridge, U.~K.:
  Cambridge University Press)

\bibitem[{Rampp {et~al.}(1998)Rampp, M\"uller, \& Ruffert}]{rampp_98_a}
Rampp, M., M\"uller, E., \& Ruffert, M. 1998, Astron.\ Astrophys., 332, 969

\bibitem[{Saijo {et~al.}(2001)Saijo, Shibata, Baumgarte, \&
  Shapiro}]{saijo_01_a}
Saijo, M., Shibata, M., Baumgarte, T.~W., \& Shapiro, S.~L. 2001,
  Astrophys.~J., 548, 919

\bibitem[{Shapiro \& Teukolsky(1983)}]{shapiro_83_a}
Shapiro, S.~L. \& Teukolsky, S.~A. 1983, Black holes, white dwarfs, and neutron
  stars (New York, U.~S.~A.: Wiley)

\bibitem[{Shibata(2000)}]{shibata_00_a}
Shibata, M. 2000, Prog.\ Theor.\ Phys., 104

\bibitem[{Shibata {et~al.}(2000)Shibata, Baumgarte, \& Shapiro}]{shibata_00_b}
Shibata, M., Baumgarte, T.~W., \& Shapiro, S.~L. 2000, Astrophys.~J., 542, 453

\bibitem[{Smith {et~al.}(1996)Smith, Houser, \& Centrella}]{smith_96_a}
Smith, S.~C., Houser, J.~L., \& Centrella, J.~M. 1996, Astrophys.~J., 458, 236

\bibitem[{Stark \& Piran(1985)}]{stark_85_a}
Stark, R.~F. \& Piran, T. 1985, Phys.\ Rev.\ Lett., 55, 891

\bibitem[{Stergioulas(1998)}]{stergioulas_98_a}
Stergioulas, N. 1998, Living Rev.\ Relativity, 1, cited on 10 Dec 2001,
  http://www.livingreviews.org/\hspace{0 cm}Articles/\hspace{0
  cm}Volume1/\hspace{0 cm}1998-8stergio/

\bibitem[{Stergioulas \& Friedman(1995)}]{stergioulas_95_a}
Stergioulas, N. \& Friedman, J.~L. 1995, Astrophys.~J., 444, 306

\bibitem[{Stergioulas \& Friedman(1998)}]{stergioulas_98_b}
---. 1998, Astrophys.~J., 492, 301

\bibitem[{Symbalisty(1984)}]{symbalisty_84_a}
Symbalisty, E.~M.~D. 1984, Astrophys.~J., 285, 729

\bibitem[{Tassoul(1978)}]{tassoul_78_a}
Tassoul, J.-L. 1978, Theory of rotating stars (Princeton, U.~S.~A.: Princeton
  University Press)

\bibitem[{Thorne(1980)}]{thorne_80_a}
Thorne, K.~S. 1980, Rev.\ Mod.\ Phys., 52, 299

\bibitem[{Thorne(1995)}]{thorne_95_a}
Thorne, K.~S. 1995, in Snowmass 1994 Summer Study on "Particle and Nuclear
  Astrophysics and Cosmology in the next millennium", ed. E.~W. Kolb \& R.~C.
  Peccei (Singapore: World Scientific), 160--184

\bibitem[{Wheeler {et~al.}(2000)Wheeler, Yi, H\"oflich, \& Wang}]{wheeler_00_a}
Wheeler, J.~C., Yi, I., H\"oflich, P., \& Wang, L. 2000, Astrophys.~J., 537,
  810

\bibitem[{Wilson(1979)}]{wilson_79_a}
Wilson, J.~R. 1979, in Sources of gravitational radiation, ed. L.~L. Smarr
  (Cambridge, U.~K.: Cambridge University Press), 423--445

\bibitem[{Wilson {et~al.}(1996)Wilson, Mathews, \& Marronetti}]{wilson_96_a}
Wilson, J.~R., Mathews, G.~J., \& Marronetti, P. 1996, Phys.\ Rev.~D, 54, 1317

\bibitem[{Yamada \& Sato(1994{\natexlab{a}})}]{yamada_95_a}
Yamada, S. \& Sato, K. 1994{\natexlab{a}}, Astrophys.~J., 450, 245

\bibitem[{Yamada \& Sato(1994{\natexlab{b}})}]{yamada_94_a}
---. 1994{\natexlab{b}}, Astrophys.~J., 434, 268

\bibitem[{Yoshida \& Eriguchi(1999)}]{yoshida_99_a}
Yoshida, S. \& Eriguchi, Y. 1999, Astrophys.~J., 515, 414

\bibitem[{Zwerger(1995)}]{zwerger_95_a}
Zwerger, T. 1995, PhD thesis, Technische Universit\"at M\"unchen, M\"unchen,
  Germany

\bibitem[{Zwerger \& M\"uller(1997)}]{zwerger_97_a}
Zwerger, T. \& M\"uller, E. 1997, Astron.\ Astrophys., 320, 209

\end{thebibliography}

\appendix


\section{Gravitational wave extraction}
\label{sec:wave_extraction}

In the appendix we summarize the technical aspects related to the
gravitational wave extraction and the different methods we have
used. As stated in Paper~I, the use of a CF metric implies the
removal of the gravitational radiation degrees of freedom. Therefore,
in order to compute the waveforms we apply the Newtonian quadrupole
formula.

In that approximation and for an axisymmetric source in spherical
coordinates, it can be shown \citep{thorne_80_a} that the transverse
traceless gravitational field has one independent component
$ h_{\theta\theta}^{\rm TT} $, which depends only on the quadrupole
signal amplitude $ A^{\rm E2}_{20} $,
\begin{equation}
  h_{\theta\theta}^{\rm TT} = \frac{1}{r} A^{\rm E2}_{20} (t - r)
  T^{{\rm E2}\,20}_{\theta\theta}.
  \label{eq:h_quadrupole_expansion}
\end{equation}
As the $ l = 2 $, $ m = 0 $ spherical tensor harmonic
$ T^{{\rm E2}\,20}_{\theta\theta} $ is defined according to
\begin{equation}
  T^{{\rm E2}\, 20}_{\theta\theta} = \frac{1}{8} \sqrt{\frac{15}{\pi}}
  \sin^2 \theta,
  \label{eq:quadrupole_tensor_harmonic}
\end{equation}
the quadrupole radiation field $ h_{\theta \theta}^{\rm TT} $ is given
by
\begin{equation}
  h_{\theta \theta}^{\rm TT} = \frac{1}{8} \sqrt{\frac{15}{\pi}}
  \sin^2 \theta \frac{A_{20}^{\rm E2}}{r}.
  \label{eq:h_tt}
\end{equation}


\subsection{Different formulations of the quadrupole formula}
\label{subsec:quadrupole_formulas}

The amplitude $ A_{20}^{\rm E2} $ in Eq.~(\ref{eq:h_tt}) is the second
time derivative of the mass quadrupole moment of the source,
\begin{equation}
  A_{20}^{\rm E2} =  \frac{{\rm d}^2}{{\rm d}t^2}
  \left( k \int \rho \left( \frac{3}{2} z^2 -
  \frac{1}{2} \right) r^4 {\rm d}r {\rm d}z \right),
\label{eq:standard_quadrupole_formula}
\end{equation}%
where $ z = \cos \theta $, and $ k = 16 \pi^{3/2} / \sqrt{15} $. This
formulation of the radiation field is known as the {\em standard
quadrupole formula}.

A direct numerical implementation of this formula is problematic, as
the discretization of the second time derivative causes high frequency
noise in the gravitational wave signal. The amplitude of this noise
can be larger than the total amplitude of the signal. The problem is
worsened by the fact that the density distribution in the integrand of
Eq.~(\ref{eq:standard_quadrupole_formula}) is weighted by $ r^4 $,
i.e.\ mass elements in the outer parts of the star -- where the grid
resolution is coarser -- contribute strongly to the signal.

To overcome this problem a standard practice is to replace the time
derivatives by spatial derivatives \citep{finn_89_a, blanchet_90_a}.
This procedure leads to the {\em stress formula} for the gravitational
wave amplitude:
\begin{equationarray}
  A_{20}^{\rm E2} =
  k \!\int\!\! \rho & \!\biggl(\!\!\! & v_r^2 (3 z^2 \!-\! 1) \!+\!
  v_\theta^2 (2 \!-\! 3 z^2) \!-\!
  v_\phi^2 \!-\! 6 r v_r v_\theta \sqrt{1 \!-\! z^2} \biggr. \nonumber \\
  & & \biggl. - r \frac{\partial \Phi}{\partial r}
  (3 z^2 \!-\! 1) \!+\! 3 \frac{\partial \Phi}{\partial \theta}
  \sqrt{1 \!-\! z^2} \biggr) r^2 {\rm d}r {\rm d}z,
  \label{eq:stress_formula}
\end{equationarray}%
where $ \Phi $ is the Newtonian gravitational potential. This modified
formulation of the quadrupole formula was used in our analysis.


\subsection{Ambiguity of the stress formula in the CFC approximation}
\label{subsec:stress_formula_ambiguity}

In a mathematically strict sense the quadrupole approximation of the
radiation field, Eq.~(\ref{eq:h_quadrupole_expansion}), is only
defined for a flat Minkowski spacetime where the gravitational waves
propagate as a linear perturbation. However, the multipole amplitudes
and the tensor harmonics, although defined in flat spacetime, contain
information about the wave source, which is situated in a strong field
and high velocity region. In our simulations the quadrupole moment
$ A_{20}^{\rm E2} $ is computed as a spatial integral over the
relativistic matter configuration in a curved spacetime, as the metric
$ g_{\mu \nu} $ is not Minkowskian. Therefore, our methods to compute
the radiation field are hampered by several ambiguities in the
quadrupole formula.

Whereas in the Newtonian formulation, which is based on Euclidean
geometry, the definition of the radius coordinate $ r $ is no source
of ambiguity, in the Riemannian geometry of a curved spacetime the
circumferential radius and the coordinate radius do not necessarily
coincide. In our simulations we use the isotropic radial coordinate
$ r $ in the computation of the quadrupole moment. Tests where we
replaced $ r $ by the circumferential radius in the wave extraction
produced almost identical waveforms.

\begin{figure}
  \resizebox{\hsize}{!}{\includegraphics{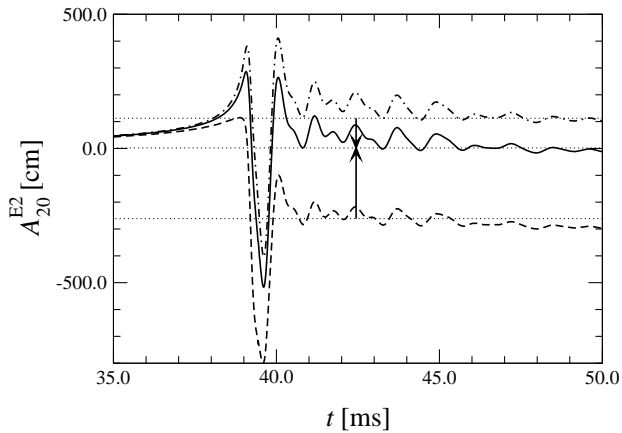}}
  \caption{Reduction of the offset in the gravitational wave amplitude
    $ A^{\rm E2}_{20} $ for model A3B2G4 with the weighting method
    applied to the stress formula. If the Newtonian potential
    $ \Phi_1 $ is used, the wave signal shows a large positive offset
    (dashed-dotted line). If the combination of metric components
    $ \Phi_2 $ is used, a large negative offset occurs (dashed
    line). The weighted sum $ \Phi_{\rm weighted} $ of these two
    signals yields a waveform with practically zero offset (solid
    line). This combined signal exhibits no high frequency noise, as
    its constituents have been obtained using the stress formula.}
  \label{fig:wave_extraction_offset_solution}
\end{figure}

The spatial derivatives of the Newtonian potential present in
Eq.~(\ref{eq:stress_formula}) need particular attention. When simply
using Poisson's equation,
\begin{equation}
  \Delta \Phi_1 = 4 \pi \rho,
  \label{eq:newtonian_potential}
\end{equation}
which we solve by an expansion into Legendre polynomials
\citep{zwerger_95_a}, the waveform shows an offset, particularly after
core bounce (see the dashed-dotted line in
Fig.~\ref{fig:wave_extraction_offset_solution}). For strong gravity,
the Newtonian potential $ \Phi_1 $ is therefore inappropriate for
describing the radiation field. Another manifestation of this problem
is a large constant offset in $ A^{\rm E2}_{20} $ for rapidly rotating
neutron stars in equilibrium, where the gravitational radiation should
vanish.

The Newtonian potential can also be computed accurately up to the
first post-Newtonian order by setting $g_{11} \approx (1 - 2 \Phi) $
equal to the conformal factor $ \phi^4 $. This leads to
\begin{equation}
  \Phi_2 = \frac{1}{2} (1 - \phi^4).
  \label{eq:alternative_expression_for_phi_newtonian}
\end{equation}
The waveforms obtained from this expression also show an offset, but
this time a negative one (see the dashed line in
Fig.~\ref{fig:wave_extraction_offset_solution}). However, combining
both expressions for $ \Phi $, Eq.~(\ref{eq:newtonian_potential}) and
Eq.~(\ref{eq:alternative_expression_for_phi_newtonian}), into
\begin{equation}
  \Phi_{\rm weighted} = \frac{\Phi_1 + a \Phi_2}{1 + a},
  \label{eq:combined_newtonian_potential}
\end{equation}
where $ a = {\rm const.} $ is a parameter with an appropriate value,
one is able to reduce the offset almost to zero in all core collapse
models (see the solid line in
Fig.~\ref{fig:wave_extraction_offset_solution}).

\end{document}